\documentclass{LMCS}

\def\dOi{11(3:10)2015}
\lmcsheading%
{\dOi}
{1--31}
{}
{}
{Nov.~17, 2014}
{Sep.~16, 2015}
{}

\ACMCCS{[{\bf Theory of computation}]: Logic---Modal and temporal
  logics; [{\bf Computer systems organization}]: Dependable and fault-tolerant systems and networks}

\usepackage{amsmath}
\usepackage{amssymb}
\usepackage{stmaryrd} 
\usepackage{wasysym}
\usepackage{mathtools} 
\usepackage{bm}
\usepackage{array}
\newcolumntype{L}{>{\qquad\labelitemi~~}l<{}}

\usepackage{subfigure}

\usepackage{verbatim}
\usepackage[hidelinks]{hyperref}
\usepackage{color}

\usepackage{cite}

\usepackage{tikz}
\usetikzlibrary{arrows,automata,positioning}
\usetikzlibrary{decorations.pathmorphing}

\tikzstyle{printing_highlight_lines}=[red,dashed]

\usepackage[nofonts]{coordinationlogic}

\newcommand{\deps}{\mathit{deps}}
\newcommand{\branches}{\mathit{branches}}
\newcommand{\paths}{\mathit{paths}}
\newcommand{\strategies}{\mathit{strategies}}
\newcommand{\header}{\mathit{header}}
\newcommand{\consistent}{\mathit{consistent}}
\newcommand{\unsatii}{\mathit{unsat}_\mathit{dist}}
\newcommand{\unsatnf}{\mathit{unsat}_\mathit{fault}}

\newcommand{\partition}{\mathit{partition}}

\newcommand{\operational}{\mathit{operational}}
\newcommand{\consensus}{\mathit{consensus}}
\newcommand{\correctval}{\mathit{correctval}}

\newcommand{\uniform}{\mathit{uniform}}

\newcommand{\Prev}{\mathit{Prev}}

\newcommand{\btrue}{\top}
\newcommand{\bfalse}{\perp}

\newcommand{\eclpuntil}   [3]{#2\;\ltluntil_{#1}\;\kern-.1em #3}

\newcommand{\modelsltl}{\vDash_\text{\rm LTL}}
\newcommand{\nmodelsltl}{\nvDash_\text{\rm LTL}}
\newcommand{\modelsqptl}{\vDash_\text{\rm QPTL}}

\newcommand{\modelssos}{\vDash_\text{\rm S1S}}
\newcommand{\nmodelssos}{\nvDash_\text{\rm S1S}}
\newcommand{\modelswsos}{\vDash_\text{\rm WS1S}}

\newcommand{\modelsqbf}{\vDash_\text{\rm QBF}}
\newcommand{\nmodelsqbf}{\nvDash_\text{\rm QBF}}

\newcommand{\sostransformer}{\mathit{s1s}}
\newcommand{\wsostransformer}{\mathit{ws1s}}
\newcommand{\qbftransformer}{\mathit{qbf}}

\theoremstyle{plain}\newtheorem{theorem}[thm]{Theorem}
\theoremstyle{plain}\newtheorem{lemma}[thm]{Lemma}
\theoremstyle{plain}\newtheorem{proposition}[thm]{Proposition}
\theoremstyle{plain}\newtheorem{corollary}[thm]{Corollary}
\theoremstyle{definition}\newtheorem{definition}{Definition}[section]

\begin{document}

\title[Detecting Unrealizability of Distributed Fault-tolerant Systems]{Detecting Unrealizability of Distributed Fault-tolerant Systems}

\author[Finkbeiner and Tentrup]{Bernd Finkbeiner}	
\address{Saarland University, 66123 Saarbr\"ucken}	
\email{\{finkbeiner,tentrup\}@cs.uni-saarland.de}  

\author[]{Leander Tentrup}	
\address{\vspace{-18 pt}}	





\begin{abstract}
  \noindent Writing formal specifications for distributed systems is difficult.
Even simple consistency requirements often turn out to be unrealizable because of the complicated information flow in the distributed system: not all information is available in every component, and information transmitted from other components may arrive with a delay or not at all, especially in the presence of faults.
The problem of checking the distributed realizability of a temporal specification is, in general, undecidable.
Semi-algorithms for synthesis, such as bounded synthesis, are only useful in the positive case, where they construct an implementation for a realizable specification, but not in the negative case: if the specification is unrealizable, the search for the implementation never terminates.
In this paper, we introduce \emph{counterexamples to distributed realizability} and present a method for the detection of such counterexamples for specifications given in linear-time temporal logic (LTL).
A counterexample consists of a set of paths, each representing a different sequence of inputs from the environment, such that, no matter how the components are implemented, the specification is violated on \emph{at least one} of these paths.
We present a method for finding such counterexamples both for the classic distributed realizability problem and for the fault-tolerant realizability problem.
Our method considers, incrementally, larger and larger sets of paths until a counterexample is found.
For safety specifications in weakly ordered architectures we obtain a decision procedure, while counterexamples for full LTL and arbitrary architectures may consist of infinitely many paths.
Experimental results, obtained with a QBF-based prototype implementation, show that our method finds simple errors very quickly, and even problems with high combinatorial complexity, like the Byzantine Generals' Problem, are tractable.
\end{abstract}

\maketitle

\section{Introduction}

The goal of program synthesis, and systems engineering in general, is to build systems that satisfy a given specification. Sometimes, however, 
this goal is unattainable, because the conditions of the specification are \emph{impossible} to satisfy in an implementation. A textbook example for such a case is the \emph{Byzantine Generals' Problem}, introduced in the early 1980s by Lamport et al.~\cite{DBLP:journals/toplas/LamportSP82}. Three generals of the Byzantine army, consisting of one commander and two lieutenants, need to agree on whether they should ``attack'' or ``retreat.''  For this purpose, the commander sends an order to the lieutenants, and all generals then exchange messages with each other, reporting, for example, to one general which messages they have received from the other general. The problem is that one of the generals is a traitor and can therefore not be assumed to tell the truth: the tale of the Byzantine generals is, after all, just an illustration for the problem of achieving fault tolerance in distributed operating systems, where we would like to achieve consensus even if a certain subset of the nodes fail. 
Of course, we cannot expect the traitor to agree with the loyal generals, but we might still expect a loyal lieutenant to agree with the order issued by a loyal commander, and two loyal lieutenants to reach a consensus in case the commander is the traitor. This specification is, however, unrealizable in the setting of the three generals (and, more generally, in all settings where at least a third of the nodes are faulty).

Detecting unrealizable specifications is of great value because it avoids spending implementation effort on specifications that are impossible to satisfy. If the system consists of a single process, then unrealizable specifications can be detected with \emph{synthesis} algorithms, which detect unrealizability as a byproduct of attempting to construct an implementation. For distributed systems, the problem is more complicated: in order to show that there is no way for the three generals to achieve consensus, we need to argue about the knowledge of each general. The key observation in the Byzantine Generals' Problem is that the loyal generals have no way of knowing who, among the other two generals, is the traitor and who is the second loyal general. For example, the situation where the commander is the traitor and orders one lieutenant to ``attack'' and the other to ``retreat'' is \emph{indistinguishable}, from the point of view of the loyal lieutenant who is ordered to attack, from the situation where the commander is loyal and orders both lieutenants to attack, while the traitor claims to have received a ``retreat'' order. Since the specification requires the lieutenant to act differently (agree with the other lieutenant vs.\ agree with the commander) in the two indistinguishable situations, we reach a contradiction. 

Since realizability for distributed systems is in general an undecidable problem~\cite{DBLP:conf/focs/PnueliR90}, the only available decision procedures are limited to special cases, such as pipeline and ring architectures~\cite{DBLP:conf/lics/KupfermanV01,DBLP:conf/lics/FinkbeinerS05}. There are semi-algorithms for distributed synthesis, such as \emph{bounded synthesis}~\cite{journals/sttt/FinkbeinerS13}, but the focus is on the search for implementations rather than on the search for inconsistencies: if an implementation exists, the semi-algorithm terminates with such an implementation, otherwise it runs forever.
In this paper, we take the opposite approach and study \emph{counterexamples to realizability}. Intuitively, a counterexample collects a sufficient number of scenarios such that, no matter what the implementation does, an error will occur in \emph{at least one} of the chosen scenarios. As specifications, we consider formulas of linear-time temporal logic (LTL). It is straightforward to encode the Byzantine Generals' Problem in LTL\@.
Another interesting example is the famous CAP Theorem, a fundamental result in the theory of distributed computation conjectured by Brewer~\cite{DBLP:conf/podc/Brewer00}. The CAP Theorem states that it is impossible to design a distributed system that provides Consistency, Availability, and Partition tolerance (CAP) simultaneously.
We assume there is a fixed number $n$ of nodes, that every node implements the same service, and that there are direct communication links between all nodes.
We use the variables $\text{req}_i$ and $\text{out}_i$ to denote input and output of node $i$, respectively.
The consistency and availability requirements can then be encoded as the LTL formulas $\bigwedge_{1 \leq i < n} (\text{out}_i \leftrightarrow \text{out}_{i+1})$ and $(\bigvee_{1 \leq i \leq n} \text{req}_i) \leftrightarrow (\eclevtl \bigvee_{1 \leq i \leq n} \text{out}_i)$.
The partition tolerance is modeled in a way that there is always at most one node partitioned from the rest of the system, i.e., we have $n$ different fault-tolerance scenarios where in every scenario all communication links to one node are faulty.

In both examples, a finite set of input sequences suffices to force the system into violating the specification on at least one of the input sequences.
In this paper, we present an efficient method for finding such counterexamples.
It turns out that searching for counterexamples is much easier than the classic synthesis approach of establishing unrealizability by the non-existence of strategy trees~\cite{DBLP:conf/focs/PnueliR90,DBLP:conf/lics/KupfermanV01,DBLP:conf/lics/FinkbeinerS05}. The difficulty in synthesis is to enforce the consistency condition that the strategy of a process must act the same way in all situations the process cannot distinguish.
On the strategy trees, this consistency condition is not an $\omega$-regular (or even decidable) property. 
When analyzing a counterexample, on the other hand, we only check consistency on a specific set of sequences, not on a full tree. This restricted consistency condition is an $\omega$-regular property and can, in fact, simply be expressed in LTL as part of the temporal specification.
Our QBF-based prototype implementation finds counterexamples for the Byzantine Generals' Problem and the CAP Theorem within just a few seconds. 

\paragraph{\bf Related Work.}
To the best of the authors' knowledge, there has been no attempt in the literature to characterize unrealizable specifications for distributed systems beyond the restricted class of architectures with decidable synthesis problems, such as pipelines and rings~\cite{DBLP:conf/lics/KupfermanV01,DBLP:conf/lics/FinkbeinerS05}.
By contrast, there is a rich literature concerning unrealizability for open systems, that is, single-process systems interacting with the environment~\cite{Church/63/Logic,DBLP:conf/icalp/AbadiLW89,conf/ictl/KupfermanV97}.
Schuppan~\cite{DBLP:journals/scp/Schuppan12} introduced the notion of unrealizable cores to identify a minimal subformula that contains the reason for unrealizability.
In robotics, there have been recent attempts to analyze unrealizable specifications~\cite{DBLP:conf/cav/RamanK11}.
The results are also focused on the reason for unrealizability, while our approach tries to determine if a specification is unrealizable.
Moreover, they only consider the simpler non-distributed synthesis of GR(1) specifications, which is a subset of LTL\@.
There are other approaches concerning unrealizable specifications in the non-distributed setting that also use counterexamples~\cite{DBLP:conf/concur/ChatterjeeHJ08,DBLP:conf/memocode/LiDS11}.
There, the system specifications are assumed to be correct and the information from the counterexamples are used to modify environment assumptions in order to make the specifications realizable.
The Byzantine Generals' Problem is often used as an illustration for knowledge-based reasoning in epistemic logics, see~\cite{DBLP:conf/podc/HalpernM84} for an early formalization.
Concerning the synthesis of fault-tolerant distributed systems, there is an approach to synthesize fault-tolerant systems in the special case of strongly connected architectures~\cite{DBLP:conf/atva/DimitrovaF09}.
A preliminary version of this paper appeared as~\cite{DBLP:conf/tacas/FinkbeinerT14}.
The present paper extends our earlier work by removing restrictions to acyclic architectures and providing a completeness result for safety specifications and weakly ordered architectures, as well as a general encoding of fault-tolerant synthesis.




\section{Distributed Realizability}

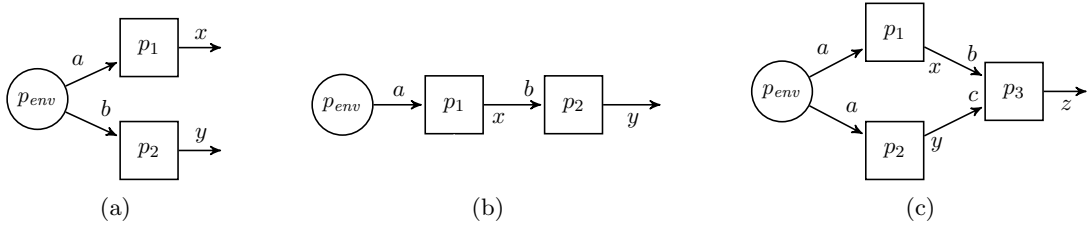
\begin{figure}[t]
\centering
\subfigure[]{
  \begin{tikzpicture}[->,>=stealth',shorten >=1pt,auto,node distance=1cm,semithick,scale=0.8,transform shape]
  
  \tikzstyle{every state}=[shape=rectangle]
  \tikzstyle{envstate}=[shape=circle,scale=0.95]
  
  \node[state,envstate]           (env) {$p_\mathit{env}$};
  \node[state,above right=0 and 1 of env] (P0)  {$p_1$};
  \node[state,below right=0 and 1 of env] (P1)  {$p_2$};
  
  \path (P0.east) edge            node  {$x$} +(right:0.75)
        (P1.east) edge            node  {$y$} +(right:0.75)
        (env) edge node {$a$} (P0)
        (env) edge node {$b$} (P1)
        ;

\end{tikzpicture}
  \label{fig:undecidable_architecture}
}\qquad%
\subfigure[]{
  \begin{tikzpicture}[->,>=stealth',shorten >=1pt,auto,node distance=1cm,semithick,scale=0.8,transform shape]
  
  \tikzstyle{every state}=[shape=rectangle]
  \tikzstyle{envstate}=[shape=circle,scale=0.95]
  
  \node[state,envstate] (env)      {$p_\mathit{env}$};
  \node[state] (P0) [right=0.85 of env] {$p_1$};
  \node[state] (P1) [right=of P0]  {$p_2$};
  
  \path (env) edge node {$a$} (P0)
        (P0) edge node[near start,swap] {$x$} node[near end] {$b$} (P1)
        (P1.east) edge node[swap] {$y$} +(1,0)
        ;
  
  \path[white] (P0.south) edge node {} +(down:0.75);

\end{tikzpicture}
  \label{fig:pipeline_architecture}
}\qquad%
\subfigure[]{
  \begin{tikzpicture}[->,>=stealth',shorten >=1pt,auto,node distance=1cm,semithick,scale=0.8,transform shape]
  \tikzstyle{every state}=[shape=rectangle]
  \tikzstyle{envstate}=[shape=circle,scale=.95]
  
  \node[state] (S1)                                {$p_3$};
  \node[state] (S2) [above left=0 and 1 of S1] {$p_1$};
  \node[state] (S3) [below left=0 and 1 of S1] {$p_2$};
  \node[state,envstate] (env) [left=2.85 of S1]     {$p_\mathit{env}$};
  
  \path (S1) edge []   node[swap] {$z$} +(1.25,0)
        (S2) edge []   node[near start,swap,xshift=5pt] {$x$} node[near end,xshift=-5pt] {$b$} (S1)
        (S3) edge []   node[near start,swap,xshift=-8pt] {$y$} node[near end,xshift=8pt,yshift=2pt] {$c$} (S1)
        (env) edge node {$a$} (S2)
        (env) edge node {$a$} (S3);
\end{tikzpicture}
  \label{fig:node_failure_example}
}
\caption[]{Example architectures}
\label{fig:architectures}
\end{figure}

A specification is \emph{realizable} if there exists an implementation that satisfies the specification. 
For distributed systems, the realizability problem is typically stated with respect to 
a specific system architecture. Figure~\ref{fig:architectures} shows some typical example architectures: an architecture consisting of \emph{independent} processes, a \emph{pipeline} architecture, and a \emph{join} architecture. The architecture describes the communication topology of the distributed system. For example, an edge from $p_1$ to $p_2$ labeled with $(x,b)$ indicates that $x$ and $b$ are shared variables between processes $p_1$ and $p_2$, where $p_1$ writes to $x$ and $p_2$ reads from $b$.
The classic \emph{distributed realizability problem} is to decide whether there exists an implementation (or \emph{strategy}) for each process in the architecture, such that the joint behavior satisfies the specification.
In this case, the distinction between the shared variables is unnecessary as the valuation of the variable that is read from is always equal to the valuation of the variable that is written to.
In this paper, we are furthermore interested in the synthesis of fault-tolerant distributed systems, where the processes and the communication between processes may become faulty.
Here, the distinction of the shared variables is used to model the different types of faults.

\paragraph{\bf LTL} \label{sec:ltl}
We use linear-time temporal logic~(LTL)~\cite{DBLP:conf/focs/Pnueli77} to express linear-time properties, that are properties $P \subseteq (2^\Sigma)^\omega$ for a finite alphabet $\Sigma$, also called \emph{atomic propositions}.
LTL consists of the temporal operators Next $\ltlnext$ and Until $\ltluntil$.
The syntax is given by the grammar
\begin{equation*} \label{eq:ltl_syntax}
\varphi \Coloneqq x \mid \neg \varphi \mid \varphi \lor \varphi \mid \eclnext \varphi \mid \ecluntil{\varphi}{\varphi} \enspace,
\end{equation*}
where $x \in \Sigma$.
We define $\varphi \land \psi$ as $\neg(\neg\varphi \lor \neg\psi)$, the Weak Until operator $\eclwuntil{\varphi}{\psi}$ as $\eclalws \varphi \lor (\ecluntil{\varphi}{\psi})$, and the Release operator $\eclrelease{\varphi}{\psi}$ as $\neg(\ecluntil{\neg\varphi}{\neg\psi})$.
We use the standard abbreviations $\true \equiv x \lor \neg x$ and $\false \equiv x \land \neg x$, for some $x \in \Sigma$, as well as $\varphi \rightarrow \psi \equiv \neg\varphi \lor \psi$, $\varphi \leftrightarrow \psi \equiv (\varphi \rightarrow \psi) \land (\psi \rightarrow \varphi)$, $\eclalws \varphi \equiv \eclrelease{\false}{\varphi}$, and $\eclevtl \varphi \equiv \ecluntil{\true}{\varphi}$.

For $i \geq 0$, the satisfaction of a path $\sigma \in (2^\Sigma)^\omega$ on position $i$ with respect to formula $\varphi$, denoted by $\sigma,i \modelsltl \varphi$, is inductively defined as\smallskip\\
\begin{tabular}{L l}
  $\sigma,i \modelsltl x$ & $:\Leftrightarrow x \in \sigma(i)$, \\[1pt]
  $\sigma,i \modelsltl \neg\varphi$ & $:\Leftrightarrow \sigma,i \nmodelsltl \varphi$, \\[1pt]
  $\sigma,i \modelsltl \varphi \lor \psi$ & $:\Leftrightarrow \sigma,i \modelsltl \varphi \text{ or } \sigma,i \modelsltl \psi$, \\[1pt]
  $\sigma,i \modelsltl \eclnext\varphi$ & $:\Leftrightarrow \sigma,i+1 \modelsltl \varphi$, and \\[1pt]
  $\sigma,i \modelsltl \ecluntil{\varphi}{\psi}$ & $:\Leftrightarrow \exists n \geq i \ldot \sigma,n \modelsltl \psi \text{ and } \forall m \in \{i,\ldots,n-1\} \ldot \sigma,m \modelsltl \varphi$ \enspace,
\end{tabular}\smallskip\\
where $x \in \Sigma$ and $\varphi, \psi$ are LTL formulas.
We say a path $\sigma \in (2^\Sigma)^\omega$ is satisfied by $\varphi$, if $\sigma,0 \modelsltl \varphi$.
The language of an LTL formula $\llbracket \varphi \rrbracket \subseteq (2^\Sigma)^\omega$ is defined as the set of paths that satisfy $\varphi$.
In~Section~\ref{sec:abstractions}, we use the syntactically restricted fragments \emph{safety LTL} and \emph{co-safety LTL}~\cite{DBLP:journals/fmsd/KupfermanV01}.
These fragments are given by the grammars
\begin{align*}
\varphi &\Coloneqq x \mid \neg x \mid \varphi \lor \varphi \mid \varphi \land \varphi \mid \eclnext \varphi \mid \eclalws \varphi \mid \eclrelease{\varphi}{\varphi}, \text{ and} \tag{safety LTL} \label{eq:safety_ltl_syntax} \\
\varphi &\Coloneqq x \mid \neg x \mid \varphi \lor \varphi \mid \varphi \land \varphi \mid \eclnext \varphi \mid \eclevtl \varphi \mid \ecluntil{\varphi}{\varphi} \enspace, \tag{co-safety LTL} \label{eq:co-safety_ltl_syntax}
\end{align*}
respectively, where $x \in \Sigma$.
Safety and co-safety formulas are dual with respect to negation.

\paragraph{\bf Coordination Logic.}
We use  Coordination logic~(CL)~\cite{DBLP:conf/csl/FinkbeinerS10} to give uniform and precise definitions of the various realizability problems of interest.
CL is a game-based extension of LTL that makes strategies---and their observations---to first class citizens of the logic.
CL divides the set of atomic propositions into strategic and coordination variables, where the latter represent observations for the strategies represented by strategic variables.
The \emph{strategy quantifier} $\eclexistsp{C}{s}$ introduces a strategy for $s$ that bases its decisions only on the history of valuations of the variables in $C$.
This allows us to pose queries on the \emph{existence} of strategies, with specific observations, within the logic.
In order to specify the realizability of distributed systems, we use the strategy quantifier $\eclexistsp{C}{s}$ to express the existence of an implementation for a process output $s$ based on input variables~$C$.

%

\paragraph{\bf CL Syntax.}
CL formulas contain two types of variables: the set $\mathcal{C}$ of \emph{input} (or \emph{coordination}) variables, and the set $\mathcal{S}$ of \emph{output} (or \emph{strategy}) variables.
In addition to the usual LTL operators Next~$\ltlnext$, Until~$\ltluntil$, and Release~$\ltlrelease$, CL has the strategy quantifier $\eclexistsp{C}{s}$, which introduces an output variable $s$ whose values must be chosen based on the history of valuations of the inputs $C$.
The syntax\footnote{This logic is called \emph{Extended} Coordination Logic in~\cite{DBLP:conf/csl/FinkbeinerS10}.} is given by the grammar
\begin{equation*} \label{eq:ecl_syntax}
\varphi \Coloneqq x \mid \neg x \mid \varphi \lor \varphi \mid \varphi \land \varphi \mid \eclnext \varphi \mid \ecluntil{\varphi}{\varphi} \mid \eclrelease{\varphi}{\varphi} \mid \eclexists{C}{s} \varphi \mid \eclforall{C}{s} \varphi \enspace ,
\end{equation*}
where $x \in \mathcal{C} \setdunion \mathcal{S}$, $C \subseteq \mathcal{C}$, and $s \in \mathcal{S}$.
Beside the standard abbreviations $\true \equiv x \lor \neg x$, $\false = x \land \neg x$, $\eclevtl \varphi \equiv \ecluntil{\true}{\varphi}$, and $\eclalws \varphi \equiv \eclrelease{\false}{\varphi}$, we use $\eclnnext{n} \varphi$ as an abbreviation of $n$ consecutive Next operators.

We denote by $\qheader$ the (possibly empty) \emph{quantification prefix} of a formula and call the remainder the \emph{body}.
For $Q \in \set{\exists,\forall}$, we use $\qheader_Q$ if the prefix contains only $Q$-quantifiers.
For the purposes of this paper, it suffices to consider the fragment CL$_\exists$ that contains only existential quantifiers.
We furthermore assume that the body is quantifier-free, i.e., that the formulas are in \emph{prenex normal form (PNF)}.

\paragraph{\bf Examples.}
We demonstrate how to express distributed realizability problems in CL$_\exists$ with the example architectures from Fig.~\ref{fig:architectures}.
The realizability of an LTL formula $\psi_1$ in the architecture from Fig.~\ref{fig:undecidable_architecture} is expressed by the CL$_\exists$ formula 
\begin{equation} \label{eq:undecidable_architecture}
\eclexists{\set{a}}{x} \eclexists{\set{b}}{y} \psi_1 \enspace .
\end{equation}
Interprocess communication via a shared variable $b$, as in the pipeline architecture from Fig.~\ref{fig:pipeline_architecture}, is expressed by separating the information read from $b$ from the output written to $b$.
In the following CL$_\exists$ formula we use output variable $x$ to denote the output written to $b$:
\begin{equation} \label{eq:pipeline_architecture}
\eclexists{\set{a}}{x} \eclexists{\set{b}}{y} \eclalws (b = x) \rightarrow \psi_2
\end{equation}
The LTL specification $\psi_2$ is qualified by the input-output specification $\eclalws (b = x)$, which expresses that $\psi_2$ is required to hold under the assumption that the information written to $b$ by process~$x$ is also the information read from $b$ by process $y$. This separation between sent and received information is useful to model faults that disturb the transmission. Failing processes are specified by omitting the input-output specifications that refer to the failing processes. As an example, consider the architecture in Fig.~\ref{fig:node_failure_example}. The CL$_\exists$ formula
\begin{equation} \label{eq:node_failure_example}
  \eclexists{\set{a}}{x,y} \eclexists{\set{b,c}}{z} \big(\eclalws (c = y) \rightarrow \psi_3 \big) \land \big(\eclalws (b = x) \rightarrow \psi_3 \big)
\end{equation}
specifies that there exists an implementation such that $\psi_3$ is guaranteed to hold even if process $x$ or $y$ (but not both) fails.

For a formula $\Phi$, we differentiate two types of coordination variables, \emph{external} and \emph{internal}.
A coordination variable $c \in \mathcal{C}$ is internal iff the value of $c$ is uniquely defined by the input-output specifications.
In contrast, external coordination variables provide input from the environment.
For example, the input $a$ in \eqref{eq:pipeline_architecture} is external while $b$ is internal.

\paragraph{\bf CL$\bm{_\exists}$ Semantics.} \label{sec:cl_semantics}
We give a quick definition of the CL$_\exists$ semantics for formulas in PNF and refer the reader to~\cite{DBLP:conf/csl/FinkbeinerS10} for details and for the semantics of full CL\@.
The semantics is based on 
\emph{trees} as a representation for strategies and computations.
Given a finite set of directions $\Upsilon$ and a finite set of labels $\Sigma$, a (full) $\Sigma$-\textit{labeled} $\Upsilon$-\textit{tree} $\tree$ is a pair $\langle \Upsilon^*, l \rangle$, where $l : \Upsilon^* \rightarrow \Sigma$ assigns each \textit{node} $\upsilon \in \Upsilon^*$ a label $l(\upsilon)$.
For two trees $\tree$ and $\tree'$, we define the joint valuation $\tree \oplus \tree'$ to be the widened tree with the union of both labels.
We refer to~\cite{DBLP:conf/csl/FinkbeinerS10} for a formal definition.
A path $\sigma$ in a $\Sigma$-labeled $\Upsilon$-tree $\tree$ is an $\omega$-word $\sigma_0 \sigma_1 \sigma_2 \ldots \in \Upsilon^\omega$ and the corresponding labeled path $\sigma^\tree$ is $(l(\epsilon), \sigma_0) (l(\sigma_0), \sigma_1) (l(\sigma_0 \sigma_1), \sigma_2) (l(\sigma_0 \sigma_1 \sigma_2), \sigma_3) \ldots \in (\Sigma \times \Upsilon)^\omega$.

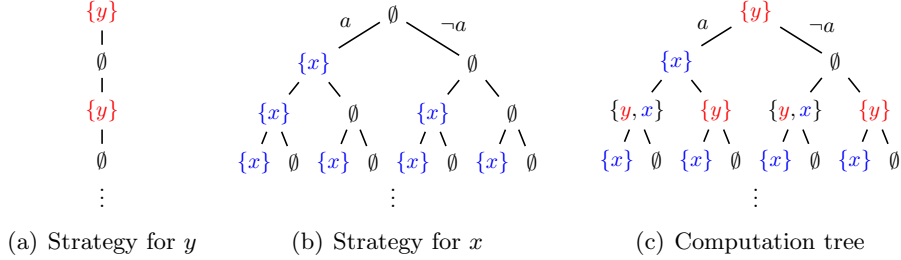
\begin{figure}[t]
  \centering
  \subfigure[Strategy for $y$] {
  \begin{tikzpicture}[auto,node distance=1cm,semithick,scale=0.75,transform shape]
      
  \tikzstyle{treenode} = [fill,circle,inner sep=0,minimum size=5pt]
  
  \node[red](root) {$\set{y}$} [level distance=25pt]
    child
    {
      node[] {$\emptyset$}
      child
      {
        node[red] {$\set{y}$}
        child
        {
          node[] {$\emptyset$}
        }
      }
    }
    ;
    
  \node[left=30pt of root] {};
  \node[right=30pt of root] {};
  
  \node[below=70pt of root] {$\vdots$};
  
  \end{tikzpicture}
  \label{fig:ecl_semantics_example_strategy_y}
}\
\subfigure[Strategy for $x$] {
  \begin{tikzpicture}[auto,node distance=1cm,semithick,scale=0.75,transform shape]
      
  \tikzstyle{treenode} = [fill,circle,inner sep=0,minimum size=5pt]
  \tikzstyle{treenodenfill} = [draw,circle,inner sep=0,minimum size=5pt]
  \tikzstyle{usepath} = [line width=1.2pt]
  
  \node[](root) {$\emptyset$} [level distance=25pt,
    level 1/.style={sibling distance=80pt},
    level 2/.style={sibling distance=40pt},
    level 3/.style={sibling distance=20pt}]
  
    child
    {
      node[blue] {$\set{x}$} 
      child
      {
        node[blue] {$\set{x}$} 
        child {node[blue] {$\set{x}$}}
        child {node[] {$\emptyset$}}
      }
      child
      {
        node[] {$\emptyset$}
        child {node[blue] {$\set{x}$}}
        child {node[] {$\emptyset$}}
      }
      edge from parent node[above left] {$a$}
    }
    child
    {
      node[] {$\emptyset$}
      child
      {
        node[blue] {$\set{x}$}
        child {node[blue] {$\set{x}$}}
        child {node[] {$\emptyset$}}
      }
      child
      {
        node[] {$\emptyset$}
        child {node[blue] {$\set{x}$}}
        child {node[] {$\emptyset$}}
      }
      edge from parent node[above right] {$\neg a$}
    }
    ;
  
  \node[below=70pt of root] {$\vdots$};
  
  \end{tikzpicture}
  \label{fig:ecl_semantics_example_strategy_x}
}\quad
\subfigure[Computation tree] {
  \begin{tikzpicture}[auto,node distance=1cm,semithick,scale=0.75,transform shape]
      
  \tikzstyle{treenode} = [fill,circle,inner sep=0,minimum size=5pt]
  \tikzstyle{treenodenfill} = [draw,circle,inner sep=0,minimum size=5pt]
  \tikzstyle{usepath} = [line width=1.2pt]
  
  \node[red](root) {$\set{y}$} [level distance=25pt,
    level 1/.style={sibling distance=80pt},
    level 2/.style={sibling distance=40pt},
    level 3/.style={sibling distance=20pt}]
  
    child
    {
      node[blue] {$\set{x}$} 
      child
      {
        node[] {$\set{{\color{red} y},{\color{blue} x}}$}
        child {node[blue] {$\set{x}$}}
        child {node[] {$\emptyset$}}
      }
      child
      {
        node[red] {$\set{y}$}
        child {node[blue] {$\set{x}$}}
        child {node[] {$\emptyset$}}
      }
      edge from parent node[above left] {$a$}
    }
    child
    {
      node[] {$\emptyset$}
      child
      {
        node[] {$\set{{\color{red} y},{\color{blue} x}}$}
        child {node[blue] {$\set{x}$}}
        child {node[] {$\emptyset$}}
      }
      child
      {
        node[red] {$\set{y}$}
        child {node[blue] {$\set{x}$}}
        child {node[] {$\emptyset$}}
      }
      edge from parent node[above right] {$\neg a$}
    }
    ;
  
  \node[below=70pt of root] {$\vdots$};
  
  \end{tikzpicture}
  \label{fig:ecl_semantics_example_computation_tree}
}

  \caption[]{In (a) and (b) we sketch example strategies for $y$ and $x$ satisfying the CL$_\exists$ formula $\eclexists{\emptyset}{y} \eclexists{\set{a}}{x} \eclalws (\eclnext x \leftrightarrow a) \land \eclalws (y \leftrightarrow \eclnext \neg y)$. In (c) we visualize the resulting computation tree on which the body (LTL) formula is evaluated.}
  \label{fig:ecl_semantics}
\end{figure}

For a strategy variable $s$ that is bound by some quantifier $\eclquantor{C}{s} \varphi$, we refer to $C$ as the \emph{scope} of $s$, denoted by $\Scope(s)$.
The meaning of a strategy variable $s$ is a \emph{strategy} or \emph{implementation} $f_s : (2^{\Scope(s)})^* \rightarrow 2^\set{s}$, i.e., a function that maps a history of valuations of input variables to a valuation of the output variable $s$.
We represent the computation of a strategy $f_s$ as the tree $\langle (2^{\Scope(s)})^*,f_s \rangle$ where $f_s$ serves as the labeling function (cf.~Fig.~\ref{fig:ecl_semantics_example_strategy_y}--\subref{fig:ecl_semantics_example_strategy_x}).
CL$_\exists$ formulas are interpreted over \emph{computation trees}, that are the joint valuations of the computations for strategies belonging to the strategy variables in $\mathcal{S}$, i.e., $\bigoplus_{s \in S} \langle (2^{\Scope(s)})^*, f_s \rangle$ (cf.~Fig.~\ref{fig:ecl_semantics}(c)).
Given a CL$_\exists$ formula $\qheader_\exists \ldot \varphi$ in prenex normal form over strategy variables $\mathcal{S}$ and coordination variables $\mathcal{C}$, the formula is satisfiable if there exists a computation tree $\tree$ (over $\mathcal{S}$), such that all paths in $\tree$ satisfy the LTL formula $\varphi$, i.e., $\forall \sigma \in (2^\mathcal{C})^\omega \ldot \sigma^\tree, 0 \vDash_\text{LTL} \varphi$.

\paragraph{\bf From Distributed Realizability to CL$\bm{_\exists}$}
We formally introduce the distributed realizability problem and show reductions from the distributed realizability problem to CL$_\exists$.
Let $V$ be a finite set of variables. An \textit{architecture} $\mathcal{A}$ is a tuple $(P,p_\mathit{env},\{I_p\}_{p \in P}, \{O_p\}_{p \in P})$, where $P$ is the set of processes and $p_\mathit{env} \notin P$ is the distinct environment process.
$I_p \subseteq V$ denotes the set of input variables for process $p$ and $O_p \subseteq V$ denotes the set of output variables for process $p$.
We denote by $I = \bigcup_{p \in P} I_p$ the set of all input variables and by $O = \bigcup_{p \in P} O_p$ the set of all output variables.
The input given by the environment is $I_\mathit{env} \coloneqq I \setminus O$, the communication variables are $I \cap O$.
While some input may be shared across processes in the case of broadcasting, the output variables of every pair of processes are assumed to be disjoint, i.e., $O_p \cap O_{p'} = \emptyset$ for all $p \neq p' \in P$.
We represent the architecture $\mathcal{A}$ by a directed graph $\mathcal{G}_\mathcal{A} = (N,E)$, where $N=P\cup P_\mathit{env}$ is the set of vertices and $E = N \times N$ the set of edges.
There is an edge between two vertices $(p,p') \in E$ if $O_p \cap I_{p'}$, i.e., there is a communication over shared variables between $p$ and $p'$.

An \textit{implementation} of a process $p$ is a function $f_p : (2^{I_p})^* \rightarrow 2^{O_p}$ which maps the history of valuation of the input variables to a subset of output variables.
We say an implementation is finite state, if it can be represented by a finite transducer.
An implementation for $\mathcal{A}$ is a set of implementations for each process.
The distributed realizability problem for architecture $\mathcal{A}$ and LTL formula $\varphi$ is to decide whether there is a finite state implementation for every process in $\mathcal{A}$ such that the system satisfies $\varphi$ against the environment, i.e., the joint behavior of the implementations satisfies $\varphi$ against all input sequences given by the environment: $\forall \sigma \in (2^I)^\omega \ldot \sigma^\tree, 0 \vDash_\text{LTL} \varphi$ where $\tree = \bigoplus_{p \in P} \langle (2^{I_p})^*, f_p \rangle$.

The distributed realizability problem is decidable for the class of 
weakly ordered architectures~\cite{DBLP:conf/lics/FinkbeinerS05}, which includes pipelines and rings.
Weakly ordered architectures are characterized by the absence of pairs of processes, called \emph{information forks}, that each have access to some information that is hidden from the other process.
Consider a tuple $(P',V',p,p')$, where $P'$ is a subset of the processes $P \cup \set{p_\mathit{env}}$, $V'$ is a subset of the variables disjoint from $I_p \cup I_{p'}$, and $p,p' \in P \smallsetminus (P' \cup \{ p_\mathit{env} \})$ are two different system processes.
Such a tuple is an information fork if $P'$ together with the edges that are
labeled with at least one variable from $V'$ forms a subgraph rooted in the
environment and there exist two nodes $q,q' \in P'$ that have edges to $p,p'$,
respectively, but are labeled with incomparable sets of variables (i.e.,
neither set is a subset of the other).
For example, the architecture in Fig.~\ref{fig:undecidable_architecture} contains the information fork $(\set{p_\mathit{env}}, \emptyset, p_1, p_2)$.
A \emph{weakly ordered architecture} is an architecture that does not contain an information fork.

\begin{definition} \label{def:induced_architecture}
  Given a CL$_\exists$ formula $\Phi = \qheader_\exists \ldot \varphi_\mathit{path} \rightarrow \varphi$, where $\varphi_\mathit{path} = \eclalws \bigwedge_{(s,c) \in R} (c = s)$ for some $R \subseteq \mathcal{C} \times \mathcal{S}$, the induced architecture $\mathcal{A}_\Phi = (P,p_\mathit{env},\{I_p\}_{p \in P}, \{O_p\}_{p \in P})$ is defined as follows.
  By abuse of notation, we use $V = \mathcal{S} \cup \mathcal{C}$ as the set of variables in the architecture and define a process $p \in P$ as a set of strategy variables $p \subseteq \mathcal{S}$.
  
  Let $P$ be the quotient of $\mathcal{S}$ according to the equivalence relation $\doteqdot{} \subseteq (\mathcal{S} \times \mathcal{S})$ with $s \doteqdot s'$ if $\Scope(s) = \Scope(s')$, i.e., we group the strategy variables with the same inputs together as a single process.
  For all $p \in P$, we define $I_p$ as $\Scope(s)$ for some $s \in p$ and $O_p$ as the union over the defined communication variables $\set{ c \mid (s,c) \in R \text{ for some } s \in p }$ and the strategy variables that are not used for communication $\set{s \mid (s,c) \notin R \text{ for all } c \in \mathcal{C}}$.
\end{definition}

\begin{theorem} \label{thm:ecl_encoding_distributed_synthesis}
The distributed realizability problem over architecture $\mathcal{A}$ and LTL formula~$\varphi$ can be encoded as CL$_\exists$ formula $\Phi = \qheader_\exists \ldot \varphi_\mathit{path} \rightarrow \varphi$ with only prenex existential quantification.
\end{theorem}
\proof
Consider an arbitrary architecture $\mathcal{A} = (P,p_\mathit{env},\{I_p\}_{p \in P}, \{O_p\}_{p \in P})$ and an LTL formula $\varphi$.
We give a CL$_\exists$ formula $\Phi$ that is satisfiable if, and only if, $\varphi$ is realizable in $\mathcal{A}$.
For variable $v \in V$, we denote the coordination variable and strategy variable used in the encoding by $c_v$ and $s_v$, respectively.
Analogously we use $C \coloneqq \{c_v \mid v \in I\}$ as the set of coordination variables and $S \coloneqq \{s_v \mid v \in O\}$ as the set of strategy variables.

For each process $p \in P$ we introduce a set of strategy variables $S_p \coloneqq \{s_v \mid v \in O_p\}$ with the scope $C_p \coloneqq \{c_v \mid v \in I_p\}$.
We take care of the input-output specifications by restricting the paths such that $\varphi_{\mathit{path}} \coloneqq \bigwedge_{v \in I \cap O} \eclalws (c_v = s_v)$ holds, i.e., we only consider paths where the valuation of the strategy variables $S$ representing the output variables is equal to the valuation of the coordination variables $I \cap O$ representing the input variables. The resulting formula $\Phi$ for processes $p_1,\ldots,p_n$ is
\begin{equation*}
\eclexistsp{C_{p_1}}{S_{p_1}} \ldots\; \eclexists{C_{p_n}}{S_{p_n}} \varphi_{\mathit{path}} \rightarrow \varphi \enspace.
\end{equation*}
The correctness of the encoding follows from the semantics of CL$_\exists$ and $\mathcal{A} = \mathcal{A}_\Phi$.
\qed

\paragraph{\bf Realizability under Byzantine faults.} 
\label{sec:multiple-architecture-realizability}
We model the occurrence of a Byzantine fault as a modification of the architecture, where a process reading from some shared variable no longer reads the output written by the other process, but instead some arbitrary input. An implementation is fault-tolerant if it works both in the original architecture and in the modified architecture.
We encode \emph{fault-tolerant realizability} as   CL$_\exists$ formulas of the general form
\begin{equation}
  \Phi = \qheader_\exists \ldot \bigwedge_{1 \leq i \leq n} \big( \varphi_{\text{path}_i} \rightarrow \varphi_i \big) \enspace ,  
\end{equation}
where the formulas $\varphi_{\text{path}_i}$ are obtained from $\varphi_{\text{path}}$ by omitting constraints $\eclalws (c = s)$, which indicates that there is a Byzantine fault on the shared variable $s$: 
the coordination variable $c$ can deviate arbitrarily from $s$.

A disadvantage of this encoding is that we can no longer partition the coordination variables into internal and external variables, because the 
coordination variable $c$ is external in the architecture where the shared variable is faulty and internal in the other architectures.
In order to retain the partitioning, we use the following satisfiability preserving transformation $\partition(\Phi)$:
For each coordination variable $c$ that is neither internal nor external, we introduce a fresh coordination variable $c^*$ that represents the external information given in the fault architectures.
Consequently, we add the condition $\eclalws (c = c^*)$ to all path specifications $\varphi_{\text{path}_i}$ where $c$ is not contained, thus, making the value $c$ deterministic in all architectures.
In the transformed formula $\partition(\Phi)$, $c$ is an internal coordination variable as its value is in every architecture uniquely determined and $c^*$ is an external coordination variable as it provides environmental input to the system.

\section{Counterexamples to Distributed Realizability} \label{sec:counterexamples}
We now introduce \emph{counterexamples to realizability}, which are actually {counterexamples to satisfiability} for the CL$_\exists$ formula encoding the realizability problem.
The satisfiability problem for a CL$_\exists$ formula in prenex form asks for an implementation for all strategy variables in the quantification prefix of the formula such that the temporal specification in the body is satisfied.

Let $\Phi=\qheader_\exists \ldot \varphi$ be a CL$_\exists$ formula in prenex form over coordination variables $\mathcal{C}$ and strategy variables $\mathcal{S}$, where the body of the formula is the LTL formula~$\varphi$.
A \emph{counterexample to satisfiability} for $\Phi$ is a (possibly infinite) set of paths $\mathcal{P} \subseteq (2^\mathcal{C})^\omega$, such that, no matter what strategies are chosen for the strategy variables in $\mathcal{S}$, there exists a path $\sigma \in \mathcal{P}$ that violates the body $\varphi$.
Formally, $\mathcal{P} \subseteq (2^\mathcal{C})^\omega$ is a counterexample to satisfiability iff, for each $s \in \mathcal{S}$ and every strategy $f_s: (2^{\Scope(s)})^* \rightarrow 2^\set{s}$, it holds that there exists a path $\sigma \in \mathcal{P}$ such that $\sigma^\tree \vDash_\text{LTL} \neg \varphi$ where $\tree = \bigoplus_{s \in \mathcal{S}} \langle (2^{\Scope(s)})^*,f_s \rangle$.
\begin{proposition} \label{eq:ecl_unsatisfiability}
A CL$_\exists$ formula $\Phi$ over coordination variables $\mathcal{C}$ and strategy variables $\mathcal{S}$
is unsatisfiable if, and only if, there exists a counterexample to satisfiability $\mathcal{P} \subseteq (2^\mathcal{C})^\omega$.
\end{proposition}
\proof
  By the semantics of CL$_\exists$ and $\mathcal{P} = (2^\mathcal{C})^\omega$.
\qed\noindent
The distributed realizability problem \emph{without} faults corresponds to CL$_\exists$ formulas of the form $\Phi = \qheader_\exists \ldot \varphi_{\text{path}} \rightarrow \varphi$, where the $\varphi_{\text{path}}$ defines the architecture $\mathcal{A}_\Phi$: there is an edge from one strategy variable to another if the input-output specification occurs in $\varphi_{\text{path}}$.
A \emph{finite} counterexample to satisfiability of $\Phi$ is a finite set of paths $\mathcal{P} \subseteq (2^{\mathcal{C}_\text{ext}})^\omega$ corresponding to external coordination variables $\mathcal{C}_\text{ext} = \set{c \mid c \text{ is an external coordination variable in } \Phi}$, such that for every implementation $\tree$ there exists a path $\sigma \in \mathcal{P}$ such that an extension $\sigma' \in (2^\mathcal{C})^\omega$ of $\sigma$ violates $\varphi$.
We say that $\sigma' \in (2^\mathcal{C})^\omega$ is an extension of $\sigma \in (2^{\mathcal{C}_\text{ext}})^\omega$ if for all $i \geq 0$, it holds that $\sigma_i = \sigma'_i \cap \mathcal{C}_\text{ext}$.
Note that the extension of $\sigma$ by the valuation of the internal coordination variables is uniquely specified by the input path $\sigma$ and the system implementation $\tree$.
\begin{proposition} \label{eq:dr_without_faults_finite_unsatisfiability}
  A CL$_\exists$ formula $\Phi = \qheader_\exists \ldot \varphi_{\text{path}} \rightarrow \varphi$ over coordination variables $\mathcal{C}$ and strategy variables $\mathcal{S}$ is unsatisfiable if there exists a finite counterexample to satisfiability $\mathcal{P} \subseteq (2^{\mathcal{C}_\text{ext}})^\omega$ of $\Phi$. \qed
\end{proposition}\noindent
As an example consider again the CL$_\exists$ formula (\ref{eq:undecidable_architecture}) $\eclexists{\set{a}}{x} \eclexists{\set{b}}{y} \psi_1$,
corresponding to the architecture from Fig.~\ref{fig:undecidable_architecture} in the previous section. Let $\psi_1 \coloneqq \eclalws (\eclnext y \leftrightarrow a)$, i.e., $y$ must output the input $a$ with an one-step delay.
A simple counterexample for this formula consists of two paths $\mathcal{P}_1 \coloneqq \set{\, \emptyset^\omega, \{a\}^\omega}$ that differ in the values of $a$, but not in the values of~$b$. 
Since process $y$ cannot distinguish the two paths, but must produce different outputs, this leads to contradiction.
Consider the same formula $\psi_1$ for the pipeline architecture specified by (\ref{eq:pipeline_architecture}) $\eclexists{\set{a}}{x} \eclexists{\set{b}}{y} \eclalws (b = x) \rightarrow \psi_1$.
The formula is unsatisfiable due to the delay when forwarding the input $a$ over shared variable $b$ (cf.\ CL$_\exists$ semantics in~Section~\ref{sec:cl_semantics}, the current output of $x$ is only visible to $y$ in the next step).
$\mathcal{P}_1$ is a finite counterexample in this case, too: Given an implementation of $x$ and $y$, we extend both paths such that the input-output specification $\eclalws (b = x)$ is satisfied.

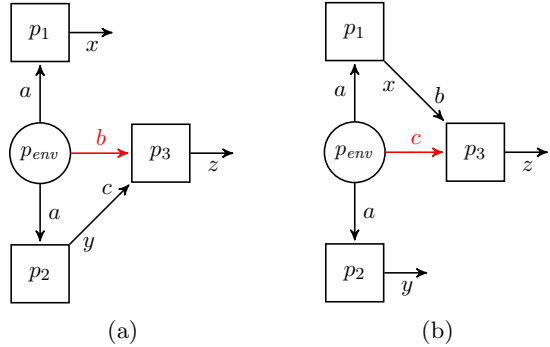
\begin{figure}[t]
\centering
\subfigure[]{
  \begin{tikzpicture}[->,>=stealth',shorten >=1pt,auto,node distance=1cm,semithick,scale=0.8,transform shape]
  \tikzstyle{every state}=[shape=rectangle]
  \tikzstyle{envstate}=[shape=circle,scale=.95]
  
  \node[state,envstate] (env) []   {$p_\mathit{env}$};
  \node[state] (S1) [right=of env] {$p_3$};
  \node[state] (S2) [above=of env] {$p_1$};
  \node[state] (S3) [below=of env] {$p_2$};
  
  \path (S1) edge []   node[swap] {$z$} +(1.25,0)
        (S2) edge []   node[swap] {$x$} +(1.25,0)
        (S3) edge []   node[near start,swap,xshift=-5pt] {$y$} node[near end,xshift=2pt,yshift=-2pt] {$c$} (S1)
        (env) edge node {$a$} (S2)
        (env) edge node {$a$} (S3)
        (env) edge[red] node {$b$} (S1);
\end{tikzpicture}
  \label{fig:node_failure_example_left}
}\qquad%
\subfigure[]{
  \begin{tikzpicture}[->,>=stealth',shorten >=1pt,auto,node distance=1cm,semithick,scale=0.8,transform shape]
  \tikzstyle{every state}=[shape=rectangle]
  \tikzstyle{envstate}=[shape=circle,scale=.95]
  
  \node[state,envstate] (env) []   {$p_\mathit{env}$};
  \node[state] (S1) [right=of env] {$p_3$};
  \node[state] (S2) [above=of env] {$p_1$};
  \node[state] (S3) [below=of env] {$p_2$};
  
  \path (S1) edge []   node[swap] {$z$} +(1.25,0)
        (S2) edge []   node[near start,swap,xshift=2pt,yshift=2pt] {$x$} node[near end,xshift=-2pt,yshift=-2pt] {$b$} (S1)
        (S3) edge []   node[swap] {$y$} +(1.25,0)
        (env) edge node {$a$} (S2)
        (env) edge node {$a$} (S3)
        (env) edge[red] node {$c$} (S1);
\end{tikzpicture}
  \label{fig:node_failure_example_right}
}
\caption[]{Visual interpretation of the fault-tolerance specification (\ref{eq:node_failure_example}): There exists a strategy for $x$, $y$, and $z$ such that the specification is satisfied in both architectures.}
\label{fig:fault_architectures}
\end{figure}

The fault-tolerant distributed realizability problem corresponds to  CL$_\exists$ formulas of the form $\Phi = \qheader_\exists \ldot \bigwedge_{1 \leq i \leq n} \big( \varphi_{\text{path}_i} \rightarrow \varphi_i \big)$.
If $\varphi_i = \varphi$ for all $i$, the formula states that there exists an implementation such that the specification $\varphi$ should hold in all architectures induced by the path specifications $\varphi_{\text{path}_i}$.
Omitted specifications $\eclalws (b = x)$ in one of these formulas $\varphi_{\text{path}_i}$ represent an arbitrary communication error at input $b$ in architecture $i$.
In this case, a counterexample identifies for each implementation one of these architectures where a contradiction occurs.
A \emph{finite} counterexample to satisfiability of $\Phi$ are $n$ finite sets of paths $\mathcal{P}_i \subseteq (2^{\mathcal{C}_\text{ext}^i})^\omega$ each corresponding to external coordination variables $\mathcal{C}_\text{ext}^i$ in the respective architecture $i$, such that for any implementation $\tree$ there exists an architecture $j$ and a path $\sigma \in \mathcal{P}_j$ such that an extension $\sigma' \in (2^\mathcal{C})^\omega$ of $\sigma$ violates $\varphi_j$.
\begin{proposition} \label{eq:dr_with_faults_finite_unsatisfiability}
  A CL$_\exists$ formula $\Phi = \qheader_\exists \ldot \bigwedge_{1 \leq i \leq n} \big( \varphi_{\text{path}_i} \rightarrow \varphi_i \big)$ over coordination variables $\mathcal{C}$ and strategy variables $\mathcal{S}$ is unsatisfiable if there exists a finite counterexample to satisfiability of $\Phi$. \qed
\end{proposition}\noindent
A counterexample for the CL$_\exists$ specification (\ref{eq:node_failure_example})
\begin{equation*}
  \eclexists{\set{a}}{x,y} \eclexists{\set{b,c}}{z} \big(\eclalws (c = y) \rightarrow \psi_3 \big) \land \big(\eclalws (b = x) \rightarrow \psi_3 \big)
\end{equation*}
introduces paths for inputs as well as for every faulty node such that some paths model the exact input-output specification and other paths model the arbitrary node failures.
The node that reads from a shared variable can, in contrast to incomplete information, react differently on the given paths, but the reaction must be consistent regarding its observations on all paths.
Consider for example the specification $\psi_3 \coloneqq (\eclnnext{2} z \leftrightarrow a)$ for the CL formula in~(\ref{eq:node_failure_example}), that is, process $z$ should output the input $a$ of nodes $x$ and $y$.
In both architectures, depicted in Fig.~\ref{fig:fault_architectures}, we introduce additional paths for the coordination variable that is omitted in the input-output specification, i.e., $b$ and $c$ for the first and second conjunct, respectively.
Process~$z$ cannot tell which of its inputs come from a faulty node.
Since $z$ must produce the same output on two paths it cannot distinguish, the implementation of $z$ contradicts the specification in either architecture.

\section{Computing Finite Counterexamples} \label{sec:computing_counterexamples}

We encode the existence of finite counterexamples to realizability as a formula of \emph{quantified propositional temporal logic (QPTL)}~\cite{DBLP:conf/lics/KestenP95}. 
QPTL extends LTL with a \emph{path quantifier} $\exists p$, where a path $\sigma \in 2^\Sigma$ satisfies $\exists p \ldot \varphi$ at position $i \geq 0$, denoted by $\sigma,i \modelsqptl \exists p \ldot \varphi$, if there exists a path $\sigma' \in 2^{\Sigma \cup \set{p}}$ which coincides with $\sigma$ except for the newly introduced atomic proposition $p$, such that $\sigma',i \modelsqptl \varphi$.
We define the universal path quantifier $\forall p \ldot \varphi$ as $\neg \exists p \ldot \neg \varphi$.
In the following encoding, we use the path quantifier to explicitly name the paths in the counterexample.

\paragraph{\bf Distributed Realizability.} \label{sec:algorithm_incomplete_information}

We consider first the distributed realizability problem, represented by CL$_\exists$ formula $\Phi = \qheader_\exists \ldot \varphi_{\text{path}} \rightarrow \varphi$.
Before proceeding with the general encoding, we show an example query that encodes the existence of a counterexample to satisfiability of CL$_\exists$ formula (\ref{eq:pipeline_architecture}) $\eclexists{\set{a}}{x} \eclexists{\set{b}}{y} \eclalws (b = x) \rightarrow \eclalws (\eclnext y \leftrightarrow a)$ by using two external paths, represented by the path variables $a_1$ and $a_2$.
The QPTL query is
\begin{align}
  \exists a_1, a_2 \ldot \forall x_1, x_2 \ldot \forall y_1, y_2 \ldot \exists b_1, b_2 \ldot \nonumber \\
  (\eclrelease{x_1 = x_2}{a_1 \neq a_2}) \land (\eclrelease{y_1 = y_2}{b_1 \neq b_2}) \rightarrow \label{eq:qptl_example_consustency} \\
  \bigwedge_{i \in \set{1,2}} \eclalws (b_i = x_i) \land \bigvee_{i \in \set{1,2}} \eclevtl (\eclnext y_i \nleftrightarrow a_i) \enspace.
\end{align}
This query is satisfiable, one satisfying assignment for $a_1$ and $a_2$ is $\set{a_1} \emptyset^\omega$ and $\emptyset^\omega$, respectively.
The assignments for $b_i$ is a (alpha renamed) copy of the assignment for $x_i$, satisfying the condition $\eclalws (b_i = x_i)$ for $i \in \set{1,2}$.
As already mentioned in the previous section, the value of $b_i$ is uniquely determined by the strategies.
Thus, we could optimize the query by removing $b_i$ altogether.
However, we need the distinction between reading and writing variables later in this section.

The valuations of $a_1$ and $a_2$ represent two \emph{paths} in the computation tree of the original CL$_\exists$ formula.
In the query above, the strategies for $x$ and $y$ are also evaluated on these two paths.
To make the connection between the path variables $x_1$ and $x_2$ and the strategy variable $x$, we introduce \emph{consistency} conditions between all paths considered in the query.
These consistency conditions~\eqref{eq:qptl_example_consustency} state that valuations of universal path variables are only valid if they can be generated by the corresponding strategy.
For example, the consistency condition for $x$, $(\eclrelease{x_1 = x_2}{a_1 \neq a_2})$, states that $a_1$ and $a_2$ must be different in some step $i \geq 0$ before $x_1$ and $x_2$ can be different in some step $j > i$.
Hence, according to~\eqref{eq:qptl_example_consustency}, $x_1 = x_2$ and $y_1 = y_2$ hold in the first step.
As $b_1 = b_2$ holds in the first step as well, the consistency condition for $y_1 = y_2$ extends to the second step as well.
Hence, we have two \emph{paths}, given by the assignments for $a_1$ and $a_2$, where $a_1$ and $a_2$ differ in the first position, but $y_1 = y_2$ in the second position.
Consequently, either $(\eclnext y_1 \nleftrightarrow a_1)$ or $(\eclnext y_2 \nleftrightarrow a_2)$ holds and we have obtained a counterexample to satisfiability of the CL$_\exists$ formula.

In the following, we introduce the general encoding for computing finite counterexamples of CL$_\exists$ formulas $\Phi = \qheader_\exists \ldot \varphi_{\text{path}} \rightarrow \varphi$ by bounding the number of paths regarding the \emph{external} coordination variables $\mathcal{C}_\text{ext}$.
The bound on the number of paths is given as a function $K : \mathcal{C} \rightarrow \mathbb{N}$ that maps each coordination variable to the number of branchings that should be considered for this variable.
W.l.o.g.\ we can assume that $K(c) = 0$ for every internal coordination variable $c \in \mathcal{C}_\text{int}$.
For example, for coordination variables $a$ and $b$, and  $K(a) = K(b) = 1$, we encode 4 different paths, one per possible combination for the two paths for each variable.
We fix an arbitrary strict order $\prec \,\subseteq \mathcal{C} \times \mathcal{C}$ between the coordination variables.
For a set $C \subseteq \mathcal{C}$, we identify $K(C)$ by the vector in $\mathbb{N}^{|C|}$ where the position of the value $K(c)$ for a coordination variable $c \in C$ is determined by $\prec$.
For our encoding in QPTL, we use the following helper functions:
\begin{itemize}
  \item $\deps(v)$ returns the set of external coordination variables that \emph{influence} variable $v$.
    An external coordination variable $c \in \mathcal{C}_\text{ext}$ influences variable $v$ if there is a directed path from $c$ to $v$ in $\mathcal{A}_\Phi$.
    For example in the architecture of Fig.~\ref{fig:node_failure_example}, $x$, $y$, and $z$ are influenced by $a$.
    A coordination variable is influenced by itself.
  
  \item $\branches(C,K)$ returns the set of branches belonging to coordination variables $C$.
    A branch $\pi$ is referenced by a vector in $\mathbb{N}^{|C|}$ and the set of branches is
    \begin{equation*}
      \Set{ (n_{c_1},\dots,n_{c_k}) \mid \set{c_1 \prec \dots \prec c_k} = C \text{ and } 1 \leq n_c \leq 2^{K(c)} \text{ for all } c \in C }
    \end{equation*}
  
  \item $\paths(C,K)$ and $\strategies(S,K)$ represent the set of (path) variables in the QPTL formula that belong to the variables of the CL$_\exists$ formula.
    For a variable $v \in C \cup S$ it introduces for each branch $\pi \in \branches(\deps(v),K)$ a separate variable $p^v_\pi$ that represents the variable $v$ belonging to this branch $\pi$.
    Formally, we define the sets
      \begin{align*}
        \paths(C,K) &\coloneqq \Set{p^c_\pi \mid c \in C \land \pi \in \branches(\deps(c),K)} \text{ and} \\
        \strategies(S,K) &\coloneqq \Set{p^s_\pi \mid s \in S \land \pi \in \branches(\deps(s),K)} \enspace .
      \end{align*}
  
  \item $\header(S,K)$ creates the alternating introductions of strategies and paths according to the (partial) order given the subset relation of the strategy variables on $\deps$.
    For every strategy variable $s \in S$ we introduce all paths belonging to external coordination variables $c \in \deps(s)$ prior to $s$ and avoid duplicate path introductions: 
    \begin{align*}
      \hspace{1cm}\header(S,K) \coloneqq {}
      &\exists\, \paths(\deps(s_1), K) \; \forall\, \strategies(\set{s_1}, K) \\
      &\exists\, \paths(\deps(s_2) \setminus \deps(s_1), K) \; \forall\, \strategies(\set{s_2}, K) \\[-5pt]
      & \dots \\[-7pt]
      &\exists\, \paths(\deps(s_n) \setminus \Big(\bigcup_{i=1,\dots,n-1} \deps(s_i)\Big), K) \; \forall\, \strategies(\set{s_n}, K) \\
      &\exists\, \paths(\mathcal{C}_\mathit{int}) \enspace ,
    \end{align*}
    where $s_1,\dots,s_n$ are ordered such that for all $i$, $j$ with $i < j \leq n$, either $\deps(s_i) \subseteq \deps(s_j)$ or $\deps(s_j) \nsubseteq \deps(s_i)$, i.e., either $\deps(s_i)$ is a subset of $\deps(s_j)$ or both are incomparable.
    
   \item $\consistent(S,K)$ specifies the consistency condition for the variables belonging to the strategy variables on the different branches.
    The variables $p^s_{\pi_1},\dots,p^s_{\pi_k}$ belonging to a strategy variable $s \in S$ must be equal as long as the coordination variables in the scope of $s$ on the branches $\pi_1,\dots,\pi_k$ are equal.
    This is expressible in LTL as we only consider a finite number of branches.
    In detail, the formula
    \begin{equation} \label{eq:consistency_conditon_qptl}
      \consistent(S,K) \coloneqq \bigwedge_{s \in S}
      \bigwedge_{\scriptsize\begin{array}{c}(\pi,\pi') \in \\\branches(\deps(s),K)^2\end{array}}
      \hspace{-15pt}
      \Big( \eclrelease{p^s_\pi = p^s_{\pi'}}{\big( \hspace{-12pt} \bigvee_{c \in \Scope(s)} \hspace{-12pt} p^c_\pi \neq p^c_{\pi'} \big)} \Big)
    \end{equation}
    states that for every strategy variable $s \in S$ and every pair of branches $(\pi,\pi')$ that belongs to $s$, we ensure that the valuation of $s$ on these two branches differ only if one of the variables in the scope of $s$ was different in a prior step.
   
\end{itemize}
Finally, we define the QPTL encoding for CL$_\exists$ formula $\Phi$ and function $K : \mathcal{C} \rightarrow \mathbb{N}$ as
\begin{align} \label{eq:incomplete_information_transformation}
  \unsatii(\Phi,K) \coloneqq
  \header(\mathcal{S}, K) \ldot
  \consistent(\mathcal{S},K) \rightarrow \nonumber \\
  \Big( \bigwedge_{\pi \in \branches(\mathcal{C},K)} \varphi_\text{path}(\pi) \Big) \land
  \Big( \bigvee_{\pi \in \branches(\mathcal{C},K)} \neg\varphi(\pi) \Big) \enspace,
\end{align}
where $\varphi(\pi)$ is the initialization of LTL formula $\varphi$ on branch $\pi$, that is we exchange $v$ by $p^v_{\pi'}$ for $v \in \mathcal{C} \cup \mathcal{S}$ where $\pi'$ is the subvector of $\pi$ that contains the values for coordination variables in $\deps(v)$.

\paragraph{\bf Fault-tolerant Realizability.} 
\label{sec:realizability_node_failures}

In the case of possible failures, the CL$_\exists$ formulas $\Phi$ have the more general form $\qheader_\exists \ldot\allowbreak \bigwedge_{1 \leq i \leq n} \big( \varphi_{\text{path}_i} \rightarrow \varphi_i \big)$.
In this setting, the set of coordination variables $\mathcal{C}$ is in general not partitioned into external and internal coordination variables.
In the first step, we apply the transformation $\partition(\Phi)$ given in Section~\ref{sec:multiple-architecture-realizability} in order to get a partitioning into external and internal coordination variables.
Furthermore, we generalize $\deps(v)$ to multiple architecture: An external coordination variable $c \in \mathcal{C}_\text{ext}$ influences variable $v$ if there is a directed path from $c$ to $v$ in one of the architectures.
The QPTL query for counterexamples to CL$_\exists$ formula $\Phi$ and a bound on the number of paths, given by the functions $K_1 \dots K_n : \mathcal{C} \rightarrow \mathbb{N}$, is
\begin{align} \label{eq:node_failure_transform}
  \unsatnf(\Phi,K_1,\dots,K_n) \coloneqq
  \header(\mathcal{S}, K) \ldot
  \consistent(\mathcal{S},K) \rightarrow \nonumber \\
  \bigvee_{1 \leq i \leq n}
  \Big( \bigwedge_{\pi \in \branches(\mathcal{C},K_i)} \varphi_{\text{path}_i}(\pi) \Big) \land
  \Big( \bigvee_{\pi \in \branches(\mathcal{C},K_i)} \neg\varphi_i(\pi) \Big) \enspace,
\end{align}
where $K : \mathcal{C} \rightarrow \mathbb{N}$ is defined as $K(c) \coloneqq \max_{1 \leq i \leq n} K_i(c)$ for every $c \in \mathcal{C}$.

Consider the example formula
\begin{equation*}
  \eclexists{\set{a,c}}{x} \eclexists{\set{b}}{z,y} (\eclalws ((c=y) \land (b=x)) \rightarrow \varphi) \land (\eclalws (c=y) \rightarrow \varphi')
\end{equation*}
that models the two-way pipeline architecture, depicted in Fig.~\ref{fig:fault_architectures}, where the connection between shared variables $b$ and $x$ may fail.
We apply the transformation $\partition(\Phi)$ and get the formula
\begin{equation*}
  \eclexists{\set{a,c}}{x} \eclexists{\set{b}}{z,y} \Big(\eclalws \big((c=y) \land (b=x)\big) \rightarrow \varphi \Big) \land \Big(\eclalws \big((c=y) \land (b=b^*)\big) \rightarrow \varphi'\Big) \enspace ,
\end{equation*}
where $b^*$ is a free coordination variable.
The QPTL encoding $\unsatnf(\Phi^*,K_1,K_2)$ with $K_1(a) = K_2(a) = K_1(b^*) = 0$ and $K_2(b^*) = 1$ is 
\begin{align}
  \exists a, b^*_1, b^*_2 \ldot \forall x_1, x_2, y_1, y_2, z_1, z_2 \ldot \exists b_1, b_2, c_1, c_2 \ldot \nonumber \\
  (\eclrelease{x_1 = x_2}{c_1 \neq c_2}) \land (\eclrelease{y_1 = y_2}{b_1 \neq b_2}) \land (\eclrelease{z_1 = z_2}{b_1 \neq b_2}) \nonumber \\
  \rightarrow \eclalws ( (c_1 = y_1) \land (b_1 = x_1) ) \land \neg\varphi(1) \label{eq:example_encoding_fault_tolerance_architecture_1}\\
  \lor \left( \bigwedge_{i \in \set{1,2}} \eclalws ((c_i = y_i) \land (b_i = b_i^*)) \right) \land \left( \bigvee_{i \in \set{1,2}} \neg \varphi(i) \right) \enspace. \label{eq:example_encoding_fault_tolerance_architecture_2}
\end{align}
In difference to distributed realizability, the encoding states that there must be a violation of the specification in some architecture.
Also, not every architecture is challenged with all paths, e.g., in the query~\eqref{eq:example_encoding_fault_tolerance_architecture_1} for the first architecture, we use only one path as the path specification states that the connection between $b$ and $x$ is correct, while in the second architecture~\eqref{eq:example_encoding_fault_tolerance_architecture_2} we use two paths that are generated by the external inputs $b^*_1$ and $b^*_2$.
\begin{figure}[t]
  \centering
  \subfigure[]{
  \begin{tikzpicture}[->,>=stealth',shorten >=1pt,auto,node distance=1.5cm,semithick,scale=0.75,transform shape]
    \tikzstyle{every state}=[shape=rectangle]
    
    \node[state] (S1)                {$p_1$};
    \node[state] (S2) [right=of S1]  {$p_2$};
    
    \path (S1) edge [<-] node {$a$} +(-1.25,0)
          (S1) edge [bend right,red,printing_highlight_lines] node[swap,near start] {$x$} node [swap,near end] {$b$} (S2)
          (S2) edge [bend right] node[swap,near start] {$y$} node[swap,near end] {$c$} (S1)
          (S2) edge [] node {$z$} +(1.25,0)
          ;
  \end{tikzpicture}
}\qquad%
\subfigure[]{
  \begin{tikzpicture}[->,>=stealth',shorten >=1pt,auto,node distance=1.5cm,semithick,scale=0.75,transform shape]
    \tikzstyle{every state}=[shape=rectangle]
    
    \node[state] (S1)                {$p_1$};
    \node[state] (S2) [right=of S1]  {$p_2$};
    
    \path (S1) edge [<-] node {$a$} +(-1.25,0)
          (S2) edge [<-] node[xshift=-2pt,yshift=2pt] {$b$} +(0,1)
          (S2) edge [] node[near start] {$y$} node[near end] {$c$} (S1)
          (S2) edge [] node {$z$} +(1.25,0)
          ;
    \node[below=6.5pt of S2] {};
  \end{tikzpicture}
}
  \caption[]{Visualization of the two architectures given by the fault-tolerance specification $\eclexists{\set{a,c}}{x} \eclexists{\set{b}}{z,y} (\eclalws ((c=y) \land (b=x)) \rightarrow \varphi) \land (\eclalws (c=y) \rightarrow \varphi')$. In architecture $(b)$, the link between shared variables $b$ and $x$ is faulty, hence, $b$ is an external input.}
  \label{fig:fault_cyclic_architecture}
\end{figure}
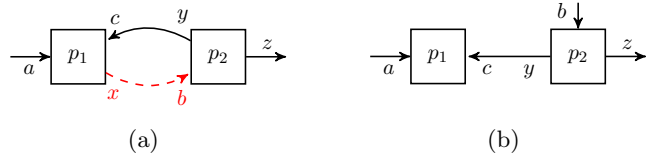

\begin{theorem}[Correctness $\unsatnf$] \label{thm:correctness_unsatnf}
  Given a CL$_\exists$ formula $\Phi = \qheader_\exists \ldot \bigwedge_{1 \leq i \leq n} \big( \varphi_{\text{path}_i} \allowbreak \rightarrow \varphi_i \big)$ over coordination variables $\mathcal{C}$ and strategy variables $\mathcal{S}$.
  $\Phi$ is unsatisfiable if there exist functions $K_1 \dots K_n : \mathcal{C} \rightarrow \mathbb{N}$ such that the QPTL formula $\unsatnf(\Phi, K_1, \dots, K_n)$ is satisfiable.
\end{theorem}
\proof
  Let $\Phi'$ be an arbitrary CL$_\exists$ formula after applying the transformation $\partition(\Phi)$ given in Section~\ref{sec:multiple-architecture-realizability} in order to recover the partitioning into external and internal coordination variables.
  Assume there exists functions $K_1 \dots K_n : \mathcal{C} \rightarrow \mathbb{N}$ such that the QPTL formula $\unsatnf(\Phi',\allowbreak K_1, \dots, K_n)$ is satisfiable.
  From the satisfiability of this QPTL formula we construct the proof that the CL$_\exists$ formula $\Phi'$ is unsatisfiable~(Proposition~\ref{eq:dr_with_faults_finite_unsatisfiability}), i.e., that for all strategies there exists a path $\sigma$ that satisfies $\bigvee_{1 \leq i \leq n} \big( \varphi_{\text{path}_i} \land \neg\varphi_i \big)$.
  
  We introduce the following auxiliary notation for this proof.
  We define $\Prev(v)$ to be the set of variables that are quantified prior to $v$.
  $\Prev_Q(v)$ denotes the set $\Prev(v)$ when only considering the existential variables ($Q=\exists$), respectively universal variables ($Q=\forall$).
  For a variable $v \in \mathcal{C} \cup \mathcal{S}$ and a branch $\pi$, we define the branching operator $v[\pi] \coloneqq p^v_{\pi'}$, where $\pi'$ is the subvector of $\pi$ that contains only values for variables in $\deps(v)$.
  $v[\pi]$ is used to relate CL variables to QPTL variables, i.e., it selects the variable in QPTL formula that \emph{belongs} to the branch $\pi$.
  Note that one QPTL variable can belong to more than one branch, for example, $v[\pi] = v[\pi']$ if the branches $\pi$ and $\pi'$ differ only in the values for variables not contained in $\deps(v)$.
  For $V \subseteq \mathcal{C} \cup \mathcal{S}$, we write $V[\pi] \coloneqq \set{ v[\pi] \mid v \in V}$ for the set of QPTL variables belonging to the CL variables $V$ on branch $\pi$.
  
  Let $s_1,\dots,s_m$ be the strategy variables $\mathcal{S}$ ordered according to the subset relation on $\deps(s)$.
  Let $\alpha(s) : (2^{\Scope(s)})^* \rightarrow 2^\set{s}$ be an arbitrary strategy for strategy variable $s \in \mathcal{S}$ and fix a set of strategies $A = \set{\alpha(s) \mid s \in \mathcal{S}}$.
  By the semantics of CL$_\exists$ we need to show that in the composition of these strategies $\bigoplus_{\alpha \in A} \alpha$ there exists a path such that the LTL formula $\bigvee_{1 \leq i \leq n} \big( \varphi_{\text{path}_i} \land \neg\varphi_i \big)$ is satisfied.
  We build the paths that satisfy the LTL formula from the branches $\Pi = \branches(\mathcal{C},K)$ encoded in the QPTL formula.
  Let $\beta(c,\pi) : (2^{\Prev_\forall(c[\pi])})^\omega \rightarrow (2^\set{c})^\omega$ be the valuation of the existential path variable in the QPTL formula belonging to coordination variable $c$ on branch $\pi \in \Pi$.
  Note that $\beta(c,\pi)$ depends on all prior universal quantified variables, especially also on universal quantified variables (representing strategy variables) on different branches.
  The strategies $\beta(c,\pi)$ for $c \in \mathcal{C}_\text{ext}$ correspond to the external coordination variables, while strategies $\beta(c,\pi)$ for $c \in \mathcal{C}_\text{int}$ correspond to the extensions defined in the finite counterexamples (Section~\ref{sec:counterexamples}).
  We view the existential quantified variables in the formula that correspond to the external coordination variables as inputs to our system strategies $A$ and get the canonical tuple of paths $\mathcal{P} = (\sigma_{\pi_1},\dots,\sigma_{\pi_k})$, one path $\sigma_\pi \subseteq (2^{\mathcal{C} \cup \mathcal{S}})^\omega$ for each branch $\pi \in \Pi$.
        
  In the construction of the paths $\mathcal{P}$, we restrict a strategy $\alpha_s \in A$ to a single branch, that is the function $\alpha_s^\omega : (2^{\Scope(s)})^\omega \rightarrow (2^\set{s})^\omega$ where $\alpha_s^\omega(\sigma) = \alpha_s(\epsilon) \concat \alpha_s(\sigma_0) \concat \alpha_s(\sigma_0 \sigma_1) \dots$.
  Let $\gamma(s,\pi) : (2^{\Prev_\exists(s[\pi])})^\omega \rightarrow (2^\set{s})^\omega$ be the strategy corresponding to the strategy variable $s \in \mathcal{S}$ in the QPTL formula for branch $\pi$.
  We have to make sure that $\gamma(s,\pi)$ is not more constrained than $\alpha_s$, i.e., the QPTL path-strategies $\gamma(s,\pi)$ must be able to handle all behaviors of $\alpha_s$.
  As $\deps(s)[\pi] \subseteq \Prev_\exists(s[\pi])$ the strategy $\gamma(s,\pi)$ subsumes the strategy $\alpha_s^\omega$, i.e., for every branch $\pi,\dots,\pi_n$ we can embed every strategy $\alpha_s$ in strategies $\gamma(s,\pi_1),\dots,\gamma(s,\pi_n)$ when only considering these branches instead of the whole computation tree.
  We build the paths $\mathcal{P}$ according to the inductive definition for every $\pi \in \Pi$ and $i \geq 1$
  \begin{align*}
    \sigma_\pi^1 = {}& \bigcup_{c \in \deps(s_1)} \beta(c,\pi)(\emptyset^\omega) , \\
    \sigma_\pi^{2i} = {}& \sigma_\pi^{2i-1} \cup \alpha_{s_i}^\omega(\sigma_\pi^{2i-1} |_{\Scope(s)}) , \text{ and}\\
    \sigma_\pi^{2i+1} = {}& \sigma_\pi^{2i} \cup \bigcup_{c \in \deps(s_i)} \beta(c,\pi)(\sigma_\pi^{2i}) \enspace,
  \end{align*}
  where $\sigma|_V$ denotes the projection of path $\sigma$ to the variables $V$.
  
  Since the $\alpha_s^\omega$ are derived from the system strategies, it holds that they satisfy the consistency condition $\consistent(\set{s},K)$~(\ref{eq:consistency_conditon_qptl}) in the QPTL encoding.
  From the satisfaction of the QPTL formula, we conclude that
  \begin{equation*}
    \bigvee_{1 \leq i \leq n} \big( \bigwedge_{\pi \in \branches(\mathcal{C},K_i)} \varphi_{\text{path}_i}(\pi) \land \bigvee_{\pi \in \branches(\mathcal{C},K_i)} \neg\varphi_i(\pi) \big)
  \end{equation*}
   is satisfied, i.e., there exists an index $i$ and branches $\Pi_i = \branches(\mathcal{C},K_i)$ such that $\bigwedge_{\pi \in \Pi_i} \varphi_{\text{path}_i}(\pi) \land \bigvee_{\pi \in \Pi_i} \neg\varphi_i(\pi)$ holds.
  From the first conjunct it follows that each path $\sigma_\pi$ from $\mathcal{P}$ corresponding to a branch $\pi \in \Pi_i$ satisfies the path specification $\varphi_{\text{path}_i}$.
  We conclude with the second conjunct that one of the branches from $\Pi$ violates the specification~$\varphi_i$.
\qed

\begin{corollary}[Correctness $\unsatii$] \label{thm:correctness_unsatii}
  Given a CL$_\exists$ formula $\Phi = \qheader_\exists \ldot \varphi_{\text{path}} \rightarrow \varphi$ over coordination variables $\mathcal{C}$ and strategy variables $\mathcal{S}$.
  $\Phi$ is unsatisfiable if there exists a function $K : \mathcal{C} \rightarrow \mathbb{N}$ such that the QPTL formula $\unsatii(\Phi,K)$ is satisfiable.
\end{corollary}

\begin{rem}[Monotonicity of $K$]
  For the distributed synthesis encoding it holds that $K$ is monotone with respect to the satisfiability of $\unsatii(\Phi,K)$, however, for the fault-tolerant synthesis this is not necessarily true because of the complex information dependencies between different architectures.
  By increasing the number of paths for a coordination variable $c$ in an architecture $i$ where $c$ is not involved in a fault, i.e., $\eclalws (c = s)$ is contained in $\varphi_{\mathit{path}_i}$, we restrict the choice of the environment in another architecture (where $c$ could be involved in a fault).
  Hence, we have the following monotonicity condition: $K_1 \dots K_n$ is monotone with respect to the satisfiability of $\unsatnf(\Phi,K_1,\dots,K_n)$ if $K_i(c)$ is only increased if $c$ is not contained in the input-output specification $\varphi_{\mathit{path}_i}$.
\end{rem}

\begin{figure}[t]
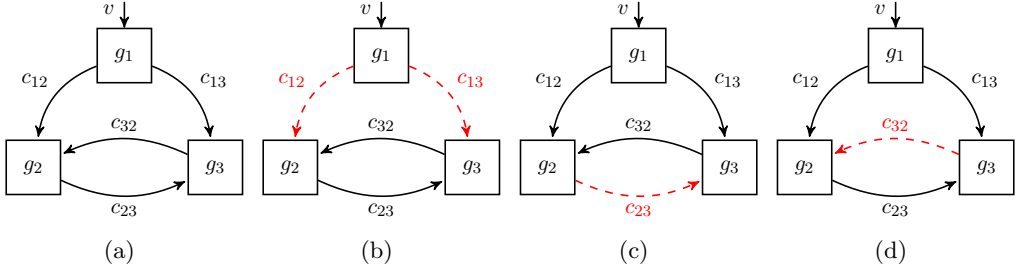

\subfigure[]{
  \begin{tikzpicture}[->,>=stealth',shorten >=1pt,auto,node distance=2cm,semithick,scale=0.75,transform shape]
    \tikzstyle{failure1}=[]
    \tikzstyle{failure2}=[]
    \tikzstyle{failure3}=[]
    \input{tikz/byzantine_architecture}
  \end{tikzpicture}
  \label{fig:byzantine_architecture_complete}
}%
\subfigure[]{
  \begin{tikzpicture}[->,>=stealth',shorten >=1pt,auto,node distance=2cm,semithick,scale=0.75,transform shape]
    \tikzstyle{failure1}=[red,printing_highlight_lines]
    \tikzstyle{failure2}=[]
    \tikzstyle{failure3}=[]
    \input{tikz/byzantine_architecture}
  \end{tikzpicture}
  \label{fig:byzantine_architecture_failure_g1}
}%
\subfigure[]{
  \begin{tikzpicture}[->,>=stealth',shorten >=1pt,auto,node distance=2cm,semithick,scale=0.75,transform shape]
    \tikzstyle{failure1}=[]
    \tikzstyle{failure2}=[red,printing_highlight_lines]
    \tikzstyle{failure3}=[]
    \input{tikz/byzantine_architecture}
  \end{tikzpicture}
  \label{fig:byzantine_architecture_failure_g2}
}%
\subfigure[]{
  \begin{tikzpicture}[->,>=stealth',shorten >=1pt,auto,node distance=2cm,semithick,scale=0.75,transform shape]
    \tikzstyle{failure1}=[]
    \tikzstyle{failure2}=[]
    \tikzstyle{failure3}=[red,printing_highlight_lines]
    \input{tikz/byzantine_architecture}
  \end{tikzpicture}
  \label{fig:byzantine_architecture_failure_g3}
}%
\caption[]{The Byzantine Generals' architecture. Figure (a) shows the architecture in cases all generals are loyal. Figures \subref{fig:byzantine_architecture_failure_g1}--\subref{fig:byzantine_architecture_failure_g3} show the possible failures, indicated by the dashed communication links.}
\label{fig:byzantine_architecture}
\end{figure}

\paragraph{\bf Example.} \label{sec:qptl_translation_example}
We consider again the Byzantine Generals' Problem with three nodes $g_1$, $g_2$, and $g_3$.
The first general is the commander who forwards the input $v$ that states whether to attack the enemy or not.
The encoding as CL$_\exists$ formula is
\begin{align} \label{eq:byzantine_generals}
  \Phi_{\text{bgp}} \coloneqq {}
  &\eclexists{\set{v}}{g_{12}, g_{13}}
  \eclexists{\set{c_{12}}}{g_{23}} 
  \eclexists{\set{c_{13}}}{g_{32}} \eclexists{\set{c_{12},c_{32}}}{g_2}
  \eclexists{\set{c_{13},c_{23}}}{g_3} \nonumber \\
  &(\operational_{2,3} \rightarrow \consensus_{2,3}) \land \bigwedge_{i \in \set{2,3}} (\operational_{1,i} \rightarrow \correctval_i) \enspace ,
\end{align}\noindent
where we use the following definitions
\begin{align*}
  \operational_{2,3} \coloneqq{}& \eclalws (c_{23} = g_{23} \land c_{32} = g_{32}) , \\
  \operational_{1,3} \coloneqq{}& \eclalws (c_{12} = g_{12} \land c_{13} = g_{13} \land c_{32} = g_{32}) , \\
  \operational_{1,2} \coloneqq{}& \eclalws (c_{12} = g_{12} \land c_{13} = g_{13} \land c_{23} = g_{23}) , \\
  \consensus_{i,j} \coloneqq{}& \eclnnext{3} (g_i = g_j) \mbox{, and} \\
  \correctval_i \coloneqq{}& v \leftrightarrow \eclnnext{3} g_i \enspace .
\end{align*}
The quantification prefix introduces the strategies for the generals $g_2$ and $g_3$, as well as the communication between the three generals as depicted in the architecture in Fig.~\ref{fig:byzantine_architecture_complete}.
Note that we omit the vote of the commander $g_1$ as it is not used in the specification.
In the temporal part, we specify which failures can occur.
The first conjunct, corresponding to Fig.~\ref{fig:byzantine_architecture_failure_g1}, states that the commander is a traitor ($\operational_{2,3}$) which implies that the other two generals have to reach a consensus whether to attack or not ($\consensus_{2,3}$).
The other two cases, depicted in Fig.~\ref{fig:byzantine_architecture_failure_g2}--\subref{fig:byzantine_architecture_failure_g3}, are symmetric and state that whenever one general is traitor the other one should agree on the decision made by the commander.
The QPTL encoding $\unsatnf(\Phi_\text{bgp}, K_1, K_2, K_3)$ is given as
\begin{align*}
  &\exists\, \paths(\set{v}, K) \ldot\;
   \forall\, \strategies(\set{g_{12},g_{13}}, K) \ldot\;
   \exists\, \paths(\set{c_{12},c_{13}}, K) \ldot \\
  &\forall\, \strategies(\set{g_{23},g_{32}}, K) \ldot\;
   \exists\, \paths(\set{c_{23},c_{32}}, K) \ldot\;
   \forall\, \strategies(\set{g_2,g_3}, K) \ldot \\  
  &\consistent(\set{g_{12},g_{13},g_{23},g_{32},g_2,g_3}, K) \rightarrow \\
    \Big( &\quad \Big( \bigwedge_{\pi \in \branches(\mathcal{C}, K_1)} \operational_{2,3}(\pi) \land \bigvee_{\pi \in \branches(\mathcal{C}, K_1)} \neg\consensus_{2,3}(\pi) \Big) \lor {}\\
    &\quad \Big( \bigwedge_{\pi \in \branches(\mathcal{C}, K_2)} \operational_{1,3}(\pi) \land \bigvee_{\pi \in \branches(\mathcal{C}, K_2)} \neg\correctval_{3}(\pi) \Big) \lor {} \\
    &\quad \Big( \bigwedge_{\pi \in \branches(\mathcal{C}, K_3)} \operational_{1,2}(\pi) \land \bigvee_{\pi \in \branches(\mathcal{C}, K_3)} \neg\correctval_{2}(\pi) \Big) \quad \Big) \enspace .
\end{align*}
By using the functions $K_i$ for $1 \leq i \leq 3$ which are defined as
\begin{align*}
  &K_1(v) = K_2(v) = K_3(v) = 1, \\
  &K_1(c_{12}) = K_1(c_{13}) = K_2(c_{23}) = K_3(c_{32}) = 1 \enspace ,
\end{align*}
and 0 otherwise, we get a satisfying QPTL instance and, hence, proved the unsatisfiability of $\Phi_\text{bgp}$.
This resembles the manual proof for the Byzantine Generals' Problem.
In every situation where one of the generals is a traitor, one has to consider the possible deviations from the loyal behavior.

\section{Completeness} \label{sec:completeness}

In practice, finite external counterexamples are often sufficient to detect unrealizable specifications. In this section, we show that our method is in fact complete for the case of safety specifications and weakly ordered architectures. 
It turns out that if one of these restrictions is dropped, i.e., if one is interested in general LTL specifications, or in architectures that are not weakly ordered, then finite counterexamples may no longer exist, and the method, therefore, becomes incomplete.


A universal safety (tree) automaton is a tuple $\mathcal{U} = (\Sigma, \Upsilon, Q, q_0, \delta)$ where $\Sigma$ denotes a finite set of labels, $\Upsilon$ denotes a finite set of directions, $Q$ is a finite set of states, $q_0 \in Q$ the designated initial state, and $\delta$ denotes the transition function.
The transition function $\delta : Q \times \Sigma \rightarrow \mathbb{B}^+_\wedge(Q \times \Upsilon)$ maps a state and a input letter to a positive and conjunctive Boolean formula over $Q \times \Upsilon$.
We allow the abbreviations $\delta(q,\sigma) = \true$ and $\delta(q,\sigma) = \false$.

A $\Sigma$-labeled $\Upsilon$-transition system is a tuple $\tree = (T,t_0,\tau,o)$ where $T$ is a finite set of states, $t_0 \in T$ is the designated initial state, $\tau : T \times \Upsilon \rightarrow T$ is the transition function, and $o : T \rightarrow \Sigma$ is the state labeling function.
A universal safety automaton accepts a $\Sigma$-labeled $\Upsilon$-transition system $\tree$ if $\tree$ has a run graph.
A run graph is a directed graph $\mathcal{G} = (G,E)$ that satisfies the following constraints:
\begin{itemize}
  \item The vertices $G \subseteq Q \times T$ are a subset of the product of $Q$ and $T$.
  \item The pair of initial states $(q_0,t_o) \in G$ is a vertex of $\mathcal{G}$.
  \item For each vertex $(q,t) \in G$, the set $\set{ (q',\upsilon) \in Q \times \Upsilon \mid ((q,t), (q',\tau(t,\upsilon))) \in E }$ satisfies $\sigma(q, o(t))$.
\end{itemize}
Since $\mathcal{U}$ is universal, the run graph on a transition system is unique.

If we start with a temporal specification instead of a universal safety automaton, we first need the following transformations.
As we only consider safety languages, we restrict the temporal specifications to \emph{syntactically safe}~\cite{DBLP:journals/fac/Sistla94} formulas, i.e., LTL formulas where the only temporal operator is the weak until $\eclwuntil{\varphi}{\psi}$. (The weak until subsumes the globally operator $\eclalws \varphi$, which is equivalent to $\eclwuntil{\varphi}{\false}$.)
\begin{proposition}[\hspace{-0.3pt}\cite{DBLP:journals/fmsd/KupfermanV01}]
  Given a syntactically safe LTL formula $\varphi$, we can construct a non-deterministic automaton $\mathcal{N}_{\neg\varphi}$ on finite words that accepts the (not necessarily minimal) good prefixes of the co-safety formula $\neg\varphi$.
  The size of $\mathcal{N}_{\neg\varphi}$ is on $2^{\mathcal{O}(|\varphi|)}$.
\end{proposition}\noindent
Using this result, we build the universal safety automaton $\mathcal{U}_\varphi$ by simulating the automaton $\mathcal{N}_{\neg\varphi}$ along each path: if each path is not a good prefix for the negation of $\varphi$, then $\varphi$ holds on every path.
\begin{proposition} \label{thm:universal-safety-automaton-for-safety-ltl}
  Given a syntactically safe LTL formula $\varphi$, we can construct a universal safety automaton $\mathcal{U}_\varphi$ with $2^{\mathcal{O}(|\varphi|)}$ states that accepts a transition system $\tree$ if, and only if, $\tree$ satisfies $\varphi$.\qed
\end{proposition}\noindent
We use the bound on the size of a transition system that is accepted by a universal safety automaton $\mathcal{U}$ in order to derive a upper bound on the number of counterexample paths needed to refute the existence of an implementation.
\begin{proposition}[\hspace{-0.3pt}\cite{journals/sttt/FinkbeinerS13}] \label{thm:safety-automaton-finite-transition-system}
If a universal safety automaton $\mathcal{U}$ with $n$ states accepts a transition system, then $\mathcal{U}$ accepts a finite transition system $\mathcal{T}$ with $(n!)^2$ states.
\end{proposition}

\begin{theorem} \label{thm:completeness_full_informed_synthesis}
  Given a syntactically safe LTL formula $\varphi$ over inputs $I$ and outputs $O$, $\varphi$ is unrealizable if, and only if, there exists a finite counterexample.
\end{theorem}
\proof
  Let $\mathcal{U}$ be the universal safety automaton for $\varphi$
  according to Proposition~\ref{thm:universal-safety-automaton-for-safety-ltl}.
  By the contraposition of Lemma~\ref{thm:safety-automaton-finite-transition-system} it holds that $\mathcal{U}$ rejects all transition systems if $\mathcal{U}$ rejects all finite transition systems with $(n!)^2$ states.
  Hence, all finite transition systems with size $(n!)^2$ have a finite counterexample path in the run of $\mathcal{U}$.
  We show that every minimal counterexample path is bounded by $d = (n!)^2 \cdot p$, where $p$ is the longest loop-free path in the safety automaton $\mathcal{U}$.
  
  Assume by contradiction that there exists a transition system $\tree$ of size $(n!)^2$, where the length of the minimal counterexample path $\sigma \in (Q \times T)^*$ exceeds $d$.
  As $d$ is the size of the product of $\mathcal{U}$ and $\tree$, there must be a repetition on $\sigma$, i.e., there exists $i < j \leq |\sigma|$ such that $\sigma[i] = \sigma[j]$.
  Thus, we can shorten the counterexample to $\sigma_{0 .. i-1} \cdot \sigma_{j .. n-1}$ violating our minimality assumption.
  
  As the length of the minimal counterexample is bounded by $d$, all minimal counterexamples are contained in the full tree of depth $d$ that has $(2^{|I|})^d$ paths.
\qed\noindent
This argumentation can be generalized to weakly ordered architectures since we have an upper bound on the size of implementations, too.
Given an architecture $\mathcal{A}$ and an implementation for each process, we call the resulting system the distributed product according to architecture $\mathcal{A}$ as the system state consist of the product of the states of the process implementations.
The meaning of the distributed product is a strategy $(2^{I_\mathit{env}}) \rightarrow 2^O$ that maps finite input sequences from the environment to the valuations of the process outputs.
\begin{proposition}[\hspace{-0.3pt}\cite{DBLP:conf/lics/FinkbeinerS05}] \label{thm:bound_distributed_realizability}
  For a weakly ordered  architecture and a realizable specification~$\varphi$, the size of the smallest implementation of every process is $n$-exponential, where $n$ is the number of processes.
\end{proposition}

\begin{theorem}
  Given a syntactically safe LTL formula $\varphi$ and a weakly ordered architecture $\mathcal{A}$, $\varphi$ is unrealizable in $\mathcal{A}$ if, and only if, there exists a finite counterexample.
\end{theorem}
\proof
   Let $\mathcal{U}$ be the universal safety automaton for $\varphi$.
  By Proposition~\ref{thm:bound_distributed_realizability} and the unrealizability of $\varphi$ in $\mathcal{A}$ it holds that there does not exist $n$-exponential strategies for the processes.
  Let $t_1,\dots,t_n$ be arbitrary $n$-exponential strategies for the $n$ processes and $\tree : (2^{I_\mathit{env}})^* \rightarrow 2^O$ be the ($n$-exponential) \emph{distributed product} according to architecture $\mathcal{A}$.
  
  We choose a bound $d$ that is greater than the size of the product of $\mathcal{U}$ and $\tree$.
  By the same argument as in the proof of Theorem~\ref{thm:completeness_full_informed_synthesis}, it follows that the length of the minimal counterexample for every such $\tree$ is bounded by $d$, hence the number of paths is bounded by $(2^{|I_\mathit{env}|})^d$.
\qed\noindent
Proposition~\ref{eq:ecl_unsatisfiability} states that the characterization of unsatisfiable formulas with counterexamples is complete.
Our method, however, searches for counterexamples involving only a bounded number of external paths. In general, this is not enough, as the following propositions show.
\begin{proposition}
  For full LTL, the unrealizability of a specification does not imply the existence of a finite counterexample.
\end{proposition}
\proof
Consider the CL$_\exists$ formula $\Phi_\text{inf} \coloneqq \eclexists{\set{x}}{y} \varphi_\text{inf}$ with temporal specification $\varphi_\text{inf} \coloneqq \eclevtl (y = x)$.
$\Phi_\text{inf}$ is unsatisfiable because for every strategy $f_y : (2^\set{x})^* \rightarrow 2^\set{y}$ it holds that $f_y$ cannot correctly guess the future value of $x$ on every path, as the formula is evaluated over the full binary tree that branches by the valuation of $x$ (cf.\ CL$_\exists$ semantics in~Section~\ref{sec:cl_semantics}).
Assume for contradiction that a finite set of paths $P \subseteq (2^\set{x})^\omega$ suffices to satisfy $\neg\varphi_\text{inf} = \eclalws (y \neq x)$ against every strategy~$f_y$.
As there are only a finite set of paths $P$, there exists a level in the full binary tree such that every node in this level has at exactly one successor path in $P$.
Choose the strategy $f_y$ that assigns the nodes in this level with one successor the labeling of the next input.
Such a finite state strategy exists because the nodes in each level are distinguishable by the strategy.
Hence, for all paths in $P$ it holds that $\eclevtl (y = x)$ and thus no path satisfies $\neg\varphi_\text{inf}$.
\qed

\begin{proposition}
  For the architecture in Fig.~\ref{fig:undecidable_architecture}, the unrealizability of a safety specification does not imply the existence of a finite counterexample.
\end{proposition}
\proof
Let $D$ be a deterministic Turing machine that implements a unary counter.
We use the encoding of $D$ into the realizability problem of a safety LTL formula $\varphi_D$ in the architecture of Fig.~\ref{fig:undecidable_architecture}~\cite{DBLP:journals/ipl/Schewe14}.
There does not exist a finite state implementation that satisfies $\varphi_D$, but for every finite set of environment input paths, only a finite number of correct configurations of $D$ can be asserted.
Hence, for every such finite set of paths, there exists an implementation that fulfills these assertions.
\qed

\section{Approximations} 
\label{sec:abstractions}

Presently available QPTL solvers were unable to handle even small instances of our problem.
In this section, we present two simplifications of the finite counterexample problem and their corresponding encodings.
We consider the problem where we restrict the counterexample paths to a finite length by using a \emph{weak monadic second order logic of one successor~(WS1S)} encoding.
Afterwards, we show a practical encoding in \emph{quantified Boolean formulas~(QBF)} that bounds the length of a counterexample path.
We only consider the more general form of CL$_\exists$ formulas modeling the fault-tolerance case, i.e., $\Phi = \qheader_\exists \ldot\allowbreak \bigwedge_{1 \leq i \leq n} \big( \varphi_{\text{path}_i} \rightarrow \varphi_i \big)$.

\paragraph{\bf From QPTL to S1S}
We give a short introduction to the monadic second-order theory of one successor (S1S) and show the equisatisfiable reduction from QPTL to S1S.
Let $V_1 = \set{x,y,\dots}$ and $V_2 = \set{X,Y,\dots}$ be finite sets of first-order and second-order variables, respectively.
A term $t$ is given by the grammar
\begin{equation*} \label{eq:s1s_terms}
  t \Coloneqq 0 \mid x \mid S(t) \enspace,
\end{equation*}
where $x$ is a first-order variable and $S$ denotes the successor function of the natural numbers.
We build formulas $\varphi$ over terms by using the grammar
\begin{equation*} \label{eq:s1s_formulas}
  \varphi \Coloneqq t \in X \mid t = t \mid \neg\varphi \mid \varphi \lor \varphi \mid \exists x\ldot \varphi \mid \exists X \ldot \varphi \enspace,
\end{equation*}
where $t$ is a term, $x$ is a first-order variable, and $X$ is a second-order variable.
We define $\forall X \ldot \varphi$ as $\neg \exists X \ldot \neg\varphi$, $x \notin X$ as $\neg (x \in X)$, and $x \neq y$ as $\neg (x = y)$.
Furthermore, we use the abbreviations $X \subseteq Y \equiv \forall z \ldot (z \in X \rightarrow z \in Y)$ and $X = Y \equiv X \subseteq Y \land Y \subseteq X$.

The semantics of S1S is defined over first-order and second-order valuations $\sigma_1 : V_1 \rightarrow \omega$ and $\sigma_2 : V_2 \rightarrow 2^\omega$, respectively.
The semantics of terms is defined as\smallskip\\
\begin{tabular}{L l}
  $[0]_{\sigma_1}$ & $= 0$,\\[1pt]
  $[x]_{\sigma_1}$ & $= \sigma_1(x)$, and\\[1pt]
  $[S(t)]_{\sigma_1}$ & $= [t]_{\sigma_1} + 1$\enspace,
\end{tabular}\smallskip\\
where $t$ is a term and $x \in V_1$.
The semantics of formulas is defined as\smallskip\\
\begin{tabular}{L l}
  $\sigma_1,\sigma_2 \modelssos t \in X$ & $:\Leftrightarrow [t]_{\sigma_1} \in \sigma_2(X)$,\\[1pt]
  $\sigma_1,\sigma_2 \modelssos t_1 = t_2$ & $:\Leftrightarrow [t_1]_{\sigma_1} = [t_2]_{\sigma_1}$,\\[1pt]
  $\sigma_1,\sigma_2 \modelssos \neg \varphi$ & $:\Leftrightarrow \sigma_1,\sigma_2 \nmodelssos \varphi$,\\[1pt]
  $\sigma_1,\sigma_2 \modelssos \varphi \lor \psi$ & $:\Leftrightarrow \sigma_1,\sigma_2 \modelssos \varphi \text{ or } \sigma_1,\sigma_2 \modelssos \psi$,\\[1pt]
  $\sigma_1,\sigma_2 \modelssos \exists x \ldot \varphi$ & $:\Leftrightarrow \exists a \in \omega \ldot \sigma'_1(y) = \begin{cases} \sigma_1(y) & \text{if } y \neq x \\ a & \text{otherwise} \end{cases} \land \sigma'_1,\sigma_2 \modelssos \varphi$, and\\[1pt]
  $\sigma_1,\sigma_2 \modelssos \exists X \ldot \varphi$ & $:\Leftrightarrow \exists A \subseteq \omega \ldot \sigma'_2(Y) = \begin{cases} \sigma_2(Y) & \text{if } Y \neq X \\ A & \text{otherwise} \end{cases} \land \sigma_1,\sigma'_2 \modelssos \varphi$ \enspace,
\end{tabular}\smallskip\\
where $t,t_1,t_2$ are terms, $\varphi,\psi$ are formulas, $x,y \in V_1$, and $X,Y \in V_2$.
A formula $\varphi$ is satisfiable, if the exists first-order and second-order valuations $\sigma_1$ and $\sigma_2$ for the free variables such that $\sigma_1,\sigma_2 \modelssos \varphi$.

It is well known that QPTL and S1S are equally expressive~\cite{DBLP:conf/icalp/SistlaVW85}.
We use the following transformation from QPTL to S1S.
\begin{lemma} \label{thm:qptl_to_s1s}
  For every QPTL formula $\varphi$ there exists an equisatisfiable S1S formula $\psi$.
\end{lemma}
\proof
We encode the (infinite) sequences in the QPTL formula as second-order valuations in the S1S formula.
For a path variable $p$, we denote by $V^p$ the corresponding second-order variable.
For a sequence $\rho \in (2^\set{p})^\omega$ and the corresponding second-order valuation $\sigma_2(V^p) \subseteq \omega$ we impose the invariant that $p \in \rho_i$ if and only if $i \in \sigma_2(V^p)$ for every $i \geq 0$.
Given a QPTL formula $\varphi$ over a finite set of atomic propositions $\Sigma$ and an arbitrary S1S term $t$ over $V_1 = \emptyset$.
We define a S1S formula $\sostransformer(\varphi,t)$ over $V_2 = \set{V^p \mid p \in \Sigma}$ such that for all $\rho \in (2^\Sigma)^\omega$ it holds that $\rho,[t]_{\sigma_1} \modelsqptl \varphi$ if and only if $\sigma_1,\sigma_2 \modelssos \sostransformer(\varphi,t)$, where $\sigma_2(V^p) = \set{i \in \omega \mid p \in \rho_i}$:\smallskip\\
\begin{tabular}{L l}
  $\sostransformer(p,t)$ & $\coloneqq t \in V^p$, for $p \in \Sigma$,\\[1pt]
  $\sostransformer(\neg\varphi, t)$ & $\coloneqq \neg \sostransformer(\varphi,t)$,\\[1pt]
  $\sostransformer(\varphi \lor \psi, t)$ & $\coloneqq \sostransformer(\varphi,t) \lor \sostransformer(\psi,t)$,\\[1pt]
  $\sostransformer(\eclnext \varphi, t)$ & $\coloneqq \sostransformer(\varphi,S(t))$,\\[1pt]
  $\sostransformer(\ecluntil{\varphi}{\psi}, t)$ & $\coloneqq \exists y \ldot (y \geq t \land \sostransformer(\psi,y) \land \forall z \ldot (t \leq z < y \rightarrow \sostransformer(\varphi,z)))$, and\\[1pt]
  $\sostransformer(\exists p \ldot \varphi, t)$ & $\coloneqq \exists V^p \ldot \sostransformer(\varphi,t)$ \enspace,
\end{tabular}\smallskip\\
where $y$ and $z$ are fresh first order variables.\qed

\paragraph{\bf Finite-length Counterexamples.}

When we restrict the counterexamples to finite length, we lose the ability to detect violations of liveness properties.
Thus, we restrict the LTL specifications $\varphi_i$ in $\Phi$ to syntactic \emph{safety} properties~\cite{DBLP:journals/fmsd/KupfermanV01}.
As these specifications are negated in the encoding, the resulting LTL specifications describe \emph{co-safety} properties~\cite{DBLP:journals/fmsd/KupfermanV01}.
A property $P \subseteq (2^\Sigma)^\omega$ is co-safety if every $\rho \in P$ has a \emph{good prefix}, that is a prefix $w \in (2^\Sigma)^*$ of $\rho$ such that for all $\rho' \in (2^\Sigma)^\omega$ it holds that $w \cdot \rho' \in P$.

In the weak version of S1S, the second-order quantification is restricted to finite sets.
While syntactically equivalent to S1S, we change the semantics $\modelssos$ by replacing the second-order quantification rule by
\begin{equation*}
  \sigma_1,\sigma_2 \modelswsos \exists X \ldot \varphi :\Leftrightarrow \exists\,\text{finite set } A \subseteq \omega \ldot \sigma'_2(Y) = \begin{cases} \sigma_2(Y) & \text{if } Y \neq X \\ A & \text{otherwise} \end{cases} \land \sigma_1,\sigma'_2 \modelswsos \varphi \enspace.
\end{equation*}
We define a S1S formula $\mathit{Fin}(X)$ that asserts that the valuation of the second-order variable $X$ is a finite set:
\begin{equation*}
  \mathit{Fin}(X) \coloneqq \exists Y \ldot (X \subseteq Y \land (\exists y \ldot y \notin Y) \land (\forall z \ldot (z \notin Y \rightarrow S(z) \notin Y))) \enspace.
\end{equation*}
We define $\wsostransformer(\varphi, i)$ as $\sostransformer(\varphi, i)$ with the difference that we interpret the result of $\wsostransformer(\varphi, i)$ as a WS1S formula and we only apply it to co-safety LTL formulas.

\begin{lemma} \label{thm:co-safety-ws1s}
Given a QPTL formula $\Phi = \mathcal{Q} \ldot \varphi$ where $\varphi$ is a co-safety LTL formula.
If $\wsostransformer(\Phi,0)$ is satisfiable then $\Phi$ is satisfiable.
\end{lemma}
\proof
Let $\wsostransformer(\Phi,0)$ be satisfiable.
By Lemma~\ref{thm:qptl_to_s1s}, it suffices to argue that $\sostransformer(\Phi,0)$ is satisfiable.
First, we can extend any existential quantification $\exists X \ldot \varphi$ to infinite quantification as the WS1S restriction to quantification over finite sets are special cases of the S1S semantics (these where $\exists X \ldot \mathit{Fin}(X) \land \varphi$ holds).
Next, we show that finite universal quantification is as powerful as infinite universal quantification.
Consider the co-safety LTL formula $\varphi$.
It holds that every satisfying assignment has a good prefix~\cite{DBLP:journals/fmsd/KupfermanV01}.
Hence, the satisfaction of an infinite word $\rho \in (2^\Sigma)^\omega$ depends only on a finite prefix.
Thus, a satisfying assignment against universal quantified finite sequences implies the satisfaction against universal quantified infinite sequences.
\qed

\noindent
The WS1S encoding of CL$_\exists$ formula $\Phi$ and functions $K_1 \dots K_n : \mathcal{C} \rightarrow \mathbb{N}$ is
\begin{align} \label{eq:node_failure_transform_wsos}
  \unsatnf^\mathit{WS1S}(\Phi,K_1,\dots,K_n) \coloneqq
  \header^\mathit{WS1S}(\mathcal{S}, K) \ldot
  \consistent^\mathit{WS1S}(\mathcal{S},K) \rightarrow \nonumber \\
  \bigvee_{1 \leq i \leq n}
  \Big( \bigwedge_{\pi \in \branches(\mathcal{C},K_i)} \varphi_{\mathit{path}_i}^\mathit{WS1S}(\pi) \Big) \land
  \Big( \bigvee_{\pi \in \branches(\mathcal{C},K_i)} \wsostransformer(\neg\varphi_i(\pi),0) \Big) \enspace,
\end{align}
where $K : \mathcal{C} \rightarrow \mathbb{N}$ is defined as $K(c) \coloneqq \max_{1 \leq i \leq n} K_i(c)$ for every $c \in \mathcal{C}$.
$\header^\mathit{WS1S}$ introduces the second-order variables $V^x$ that correspond to the path variables $x$, but is otherwise identical to the QPTL version.
Furthermore, $\varphi_\mathit{path}^\mathit{WS1S}$ uses second-order equivalence $V^s = V^c$ for $\eclalws (c \leftrightarrow v)$.
The consistency condition $\consistent^\mathit{WS1S}(\mathcal{S}, K)$ in WS1S is defined as the conjunction of
\begin{equation*}
  \bigwedge_{(\pi,\pi') \in \branches(\deps(s),K)^2} \hspace{-24pt} \Big( V^s_\pi = V^s_{\pi'} \lor \exists i \geq 0 \ldot \big( \hspace{-10pt} \bigvee_{c \in \Scope(s)} \hspace{-10pt} i \in V^c_\pi \nleftrightarrow i \in V^c_{\pi'} \big) \land \forall j \leq i \ldot j \in V^s_\pi \leftrightarrow j \in V^s_{\pi'} \Big)
\end{equation*}
for every $s \in \mathcal{S}$.
This formula ensures different reactions of a universal variable on two branches is based on different prior valuations of the dependencies (cf.~QPTL consistency condition~\eqref{eq:consistency_conditon_qptl}).

We give an example query based on CL$_\exists$ formula (\ref{eq:pipeline_architecture}) $\eclexists{\set{a}}{x} \eclexists{\set{b}}{y} \eclalws (b = x) \rightarrow \eclalws (\eclnext y \leftrightarrow a)$.
The WS1S query is
\begin{align}
  \exists A_1, A_2 \ldot \forall X_1, X_2 \ldot \forall Y_1, Y_2 \ldot \exists B_1, B_2 \ldot \nonumber \\
  \Big( X_1 = X_2 \lor \big( \exists i \geq 0 \ldot (i \in A_1 \nleftrightarrow i \in A_2) \land \forall j \leq i \ldot j \in X_1 \leftrightarrow j \in X_2 \big) \Big) \land {} \nonumber \\
  \Big( Y_1 = Y_2 \lor \big( \exists i \geq 0 \ldot (i \in B_1 \nleftrightarrow i \in B_2) \land \forall j \leq i \ldot j \in Y_1 \leftrightarrow j \in Y_2 \big) \Big) \rightarrow {} \nonumber \\
  \bigwedge_{i \in \set{1,2}} (B_i = X_i) \land \bigvee_{i \in \set{1,2}} \exists k \geq 0 \ldot (S(k) \in Y_i \nleftrightarrow k \in A_i) \enspace.
\end{align}

\begin{theorem}[Correctness $\unsatnf^\mathit{WS1S}$] \label{thm:correctness_ws1s_encoding}
  Given a CL$_\exists$ formula $\Phi = \qheader_\exists \ldot \bigwedge_{1 \leq i \leq n} \big( \varphi_{\text{path}_i} \allowbreak \rightarrow \varphi_i \big)$ over coordination variables $\mathcal{C}$ and strategy variables $\mathcal{S}$.
  Let $\varphi_i$ be a syntactically safe LTL formula for each $1 \leq i \leq n$.
  $\Phi$ is unsatisfiable if there exists functions $K_1 \dots K_n : \mathcal{C} \rightarrow \mathbb{N}$ such that the WS1S formula $\unsatnf^\mathit{WS1S}(\Phi,K_1,\dots,K_n)$ is satisfiable.
\end{theorem}
\proof
Assume $\unsatnf^\mathit{WS1S}(\Phi,K_1,\dots,K_n)$ is satisfiable.
We show that this implies the satisfiability of $\unsatnf(\Phi,K_1,\dots,K_n)$ and use Theorem~\ref{thm:correctness_unsatnf} that states the correctness of the QPTL encoding.
Lemma~\ref{thm:co-safety-ws1s} states that we only have to consider finite state sequences for the LTL specifications $\varphi_i$.
What is missing in the argumentation are the paths specifications $\varphi_{\text{path}_i}^\mathit{WS1S}$ and the consistency condition $\consistent^\mathit{WS1S}(\Phi,K)$.
The transformed path specifications are equisatisfiable under WS1S and QPTL semantics as the (existential quantified) input variables appear after the (universal quantified) strategy variables in $\header^\mathit{WS1S}(\mathcal{S},K)$.
Lastly, the consistency condition restricts the finite behavior of the universal quantified strategy variables and the infinite behavior does not affect the satisfiability of the co-safety properties.
Hence, there exists an infinite path extension for every existentially quantified second-order variable, and the formula remains satisfiable against any infinite path extension of the universally quantified second-order variables.
\qed\noindent
WS1S is supported by the Mona~\cite{DBLP:conf/tacas/HenriksenJJKPRS95} tool.
Some of our smaller instances were solved by Mona, but the Byzantine Generals' Problem failed due to memory constraints in the BDD library.

\paragraph{\bf Bounded-length Counterexamples.}
Taking the simplifications even further, we not only bound the \emph{number} of paths but also the \emph{length} of the paths by translating the problem to the satisfiability problem of \emph{quantified Boolean formulas (QBF)}.
Quantified Boolean formulas are the extension of Boolean formulas by quantification over variables.
Let $V = \set{v_1,\dots,v_n}$ be a finite set of variables.
The syntax of QBF is given by the grammar
\begin{equation*} \label{eq:qbf_syntax}
  \varphi \Coloneqq x \mid \neg\varphi \mid \varphi \lor \varphi \mid \exists x \ldot \varphi \enspace,
\end{equation*}
where $x \in V$.
We define the usual abbreviations $\land$, $\forall$, $\true$, $\false$, $\rightarrow$, and $\leftrightarrow$.
The semantics is defined over valuations $\sigma : V \rightarrow \set{0,1}$. The satisfaction of an valuation $\sigma$ is defined as\smallskip\\
\begin{tabular}{L l}
  $\sigma \modelsqbf x$ & $:\Leftrightarrow \sigma(x) = 1$,\\[1pt]
  $\sigma \modelsqbf \neg\varphi$ & $:\Leftrightarrow \sigma \nmodelsqbf \varphi$, \\[1pt]
  $\sigma \modelsqbf \varphi \lor \psi$ & $:\Leftrightarrow \sigma \modelsqbf \varphi \text{ or } \sigma \modelsqbf \psi$, and\\[1pt]
  $\sigma \modelsqbf \exists x \ldot \varphi$ & $:\Leftrightarrow \exists a \in \set{0,1} \ldot \sigma'(y) = \begin{cases} \sigma(y) & \text{if } y \neq x \\ a & \text{otherwise} \end{cases} \land \sigma' \modelsqbf \varphi$ \enspace.
\end{tabular}\smallskip\\
The encoding translates a QPTL variable $x$ to Boolean variables $x_0,\dots,x_{k-1}$, each representing one step in the system where $k$ is the length of the paths.
We build the QBF formula by \emph{unrolling} the QPTL formula $\unsatnf(\Phi,K_1,\dots,K_n)$ for $k$-steps: Each variable in the quantification prefix of the QPTL formula is transformed into $k$ Boolean variables in the QBF prefix, e.g., the 3-unrolling of $\exists x \ldot \forall y \ldot \varphi$ is $\exists x_0, x_1, x_2 \ldot \forall y_0, y_1, y_2 \ldot \varphi_\mathit{unroll}$.
The unrolling of the remaining LTL formula is given by the expansion law for Until, $\ecluntil{\varphi}{\psi} \equiv \psi \lor (\varphi \land \eclnext \ecluntil{\varphi}{\psi})$.
After the unrolling, the QBF formula is transformed into Conjunctive Normal Form~(CNF) and encoded in the QDIMACS file format, that is the standard format for QBF solvers.
Already with this encoding we could solve more examples than using the WS1S approach.

\begin{figure}
  \subfigure[]{
    \tikzstyle{used1}=[]
    \tikzstyle{used2}=[]
    \tikzstyle{used3}=[]
    \tikzstyle{used4}=[]
    \tikzstyle{unused}=[]
      \begin{tikzpicture}[node distance=0.5cm,scale=0.75,semithick,transform shape]
    
    \node (step0) {$0$};
    \node[used1,right=0.5cm of step0] (v0) {$v$};
    \node[unused,right=of v0] (g120) {$g_{12}$};
    \node[unused,right=of g120] (g130) {$g_{13}$};
    \node[unused,right=of g130] (g230) {$g_{23}$};
    \node[unused,right=of g230] (g320) {$g_{32}$};
    \node[unused,right=of g320] (g20) {$g_2$};
    \node[unused,right=of g20] (g30) {$g_3$};
    
    \node[below=1.5cm of step0] (step1) {$1$};
    \node[right=0.5cm of step1] (v1) {$v$};
    \node[right=of v1] (g121) {$g_{12}$};
    \node[right=of g121] (g131) {$g_{13}$};
    \node[unused,right=of g131] (g231) {$g_{23}$};
    \node[unused,right=of g231] (g321) {$g_{32}$};
    \node[unused,right=of g321] (g21) {$g_2$};
    \node[unused,right=of g21] (g31) {$g_3$};
    
    \node[below=1.5cm of step1] (step2) {$2$};
    \node[unused,right=0.5cm of step2] (v2) {$v$};
    \node[right=of v2] (g122) {$g_{12}$};
    \node[right=of g122] (g132) {$g_{13}$};
    \node[right=of g132] (g232) {$g_{23}$};
    \node[right=of g232] (g322) {$g_{32}$};
    \node[unused,right=of g322] (g22) {$g_2$};
    \node[unused,right=of g22] (g32) {$g_3$};
    
    \node[below=1.5cm of step2] (step3) {$3$};
    \node[unused,right=0.5cm of step3] (v3) {$v$};
    \node[unused,right=of v3] (g123) {$g_{12}$};
    \node[unused,right=of g123] (g133) {$g_{13}$};
    \node[unused,right=of g133] (g233) {$g_{23}$};
    \node[unused,right=of g233] (g323) {$g_{32}$};
    \node[used1,right=of g323] (g23) {$g_2$};
    \node[used1,right=of g23,yshift=2pt] (g33) {$g_3$};
    
    \path[unused]
          (g120) edge (g231)
          (g120) edge (g21)
          (g130) edge (g321)
          (g130) edge (g31)
          (g230) edge (g31)
          (g320) edge (g21)
          
          (g121) edge (g22)
          (g131) edge (g32)
          (g231) edge (g32)
          (g321) edge (g22)
          
          (v2) edge (g123)
          (v2) edge (g133)
          (g122) edge (g233)
          (g132) edge (g323)
          ;
    
    \path[used4]
          (v0) edge (g121)
          (v0) edge (g131)
          ;

    \path[used3]
          (v1) edge (g122)
          (v1) edge (g132)
          (g121) edge (g232)
          (g131) edge (g322)
          ;
          
    \path[used2]
          (g122) edge (g23)
          (g132) edge (g33)
          (g232) edge (g33)
          (g322) edge (g23)
          ;
    
    \draw (step0.north east) -- (step3.south east);
    
  \end{tikzpicture}
    \label{fig:qbf_encoding_optimization_dependency_graph_before}
  }%
  \subfigure[]{
    \tikzstyle{used1}=[red,circle,draw]
    \tikzstyle{used2}=[red]
    \tikzstyle{used3}=[red]
    \tikzstyle{used4}=[red]
    \tikzstyle{unused}=[black!30]
      \begin{tikzpicture}[node distance=0.5cm,scale=0.75,semithick,transform shape]
    
    \node (step0) {$0$};
    \node[used1,right=0.5cm of step0] (v0) {$v$};
    \node[unused,right=of v0] (g120) {$g_{12}$};
    \node[unused,right=of g120] (g130) {$g_{13}$};
    \node[unused,right=of g130] (g230) {$g_{23}$};
    \node[unused,right=of g230] (g320) {$g_{32}$};
    \node[unused,right=of g320] (g20) {$g_2$};
    \node[unused,right=of g20] (g30) {$g_3$};
    
    \node[below=1.5cm of step0] (step1) {$1$};
    \node[right=0.5cm of step1] (v1) {$v$};
    \node[right=of v1] (g121) {$g_{12}$};
    \node[right=of g121] (g131) {$g_{13}$};
    \node[unused,right=of g131] (g231) {$g_{23}$};
    \node[unused,right=of g231] (g321) {$g_{32}$};
    \node[unused,right=of g321] (g21) {$g_2$};
    \node[unused,right=of g21] (g31) {$g_3$};
    
    \node[below=1.5cm of step1] (step2) {$2$};
    \node[unused,right=0.5cm of step2] (v2) {$v$};
    \node[right=of v2] (g122) {$g_{12}$};
    \node[right=of g122] (g132) {$g_{13}$};
    \node[right=of g132] (g232) {$g_{23}$};
    \node[right=of g232] (g322) {$g_{32}$};
    \node[unused,right=of g322] (g22) {$g_2$};
    \node[unused,right=of g22] (g32) {$g_3$};
    
    \node[below=1.5cm of step2] (step3) {$3$};
    \node[unused,right=0.5cm of step3] (v3) {$v$};
    \node[unused,right=of v3] (g123) {$g_{12}$};
    \node[unused,right=of g123] (g133) {$g_{13}$};
    \node[unused,right=of g133] (g233) {$g_{23}$};
    \node[unused,right=of g233] (g323) {$g_{32}$};
    \node[used1,right=of g323] (g23) {$g_2$};
    \node[used1,right=of g23,yshift=2pt] (g33) {$g_3$};
    
    \path[unused]
          (g120) edge (g231)
          (g120) edge (g21)
          (g130) edge (g321)
          (g130) edge (g31)
          (g230) edge (g31)
          (g320) edge (g21)
          
          (g121) edge (g22)
          (g131) edge (g32)
          (g231) edge (g32)
          (g321) edge (g22)
          
          (v2) edge (g123)
          (v2) edge (g133)
          (g122) edge (g233)
          (g132) edge (g323)
          ;
    
    \path[used4]
          (v0) edge (g121)
          (v0) edge (g131)
          ;

    \path[used3]
          (v1) edge (g122)
          (v1) edge (g132)
          (g121) edge (g232)
          (g131) edge (g322)
          ;
          
    \path[used2]
          (g122) edge (g23)
          (g132) edge (g33)
          (g232) edge (g33)
          (g322) edge (g23)
          ;
    
    \draw (step0.north east) -- (step3.south east);
    
  \end{tikzpicture}
    \label{fig:qbf_encoding_optimization_dependency_graph_after}
  }%
  \caption[]{Example for a dependency graph of the Byzantines' Generals Problem. The graph identifies all variables that influence the variables used in the LTL specifications $\consensus_{i,j} \coloneqq \eclnnext{3} (g_i = g_j)$ and $\correctval_i \coloneqq v \leftrightarrow \eclnnext{3} g_i$ (variables with red circle). All variables that do not influence the LTL formula can be safely removed, e.g., the input $v$ in step $2$ and $3$ never reaches the strategies $g_2$ and $g_3$.}
  \label{fig:qbf_encoding_optimization_dependency_graph}
\end{figure}
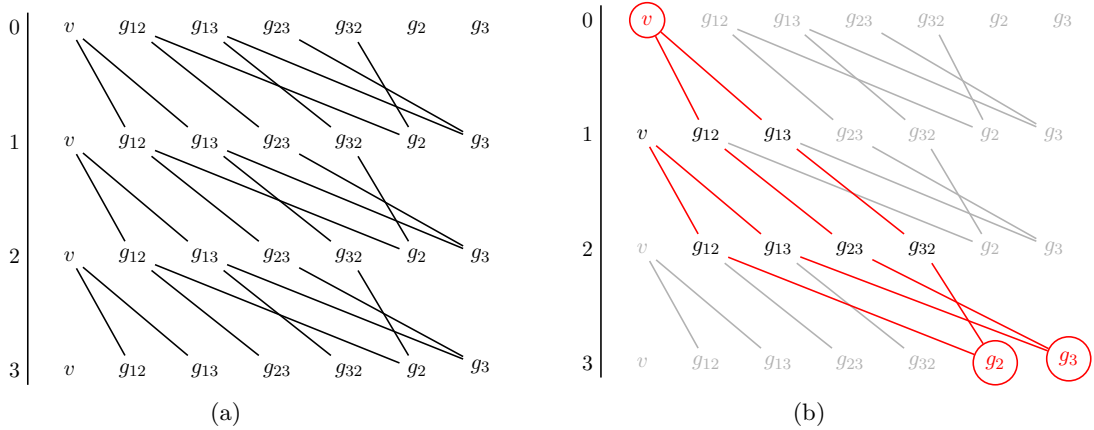

Formally, we define the transformation function $\qbftransformer(\Phi,i,k)$ that takes a QPTL formula $\Phi$ over alphabet $\Sigma$ and transforms it into the $k$-unrolling:\smallskip\\
\begin{tabular}{L l}
  $\qbftransformer(p, i, k)$ & $\coloneqq p_i$, for $p \in \Sigma$,\\[1pt]
  $\qbftransformer(\neg p, i, k)$ & $\coloneqq \neg p_i$, for $p \in \Sigma$,\\[1pt]
  $\qbftransformer(\varphi_1 \circ \varphi_2, i, k)$ & $\coloneqq \qbftransformer(\varphi_1, i, k) \circ \qbftransformer(\varphi_2, i, k)$ for $\circ \in \set{\lor,\land}$,\\[1pt]
  $\qbftransformer(\eclnext \varphi, i, k)$ & $\coloneqq \begin{cases}
    \qbftransformer(\varphi, i+1, k) & \text{ if } i < k \\
    \false & \text{ otherwise}
  \end{cases}$,\\[1pt]
  $\qbftransformer(\ecluntil{\varphi}{\psi}, i, k)$ & $\coloneqq \begin{cases}
    \qbftransformer(\psi, i, k) \lor (\qbftransformer(\varphi, i, k) \land \qbftransformer(\ecluntil{\varphi}{\psi}, i+1, k)) & \text{ if } i < k \\
    \false & \text{ otherwise}
  \end{cases}$,\\[1pt]
  $\qbftransformer(Q\, p \ldot \varphi, i, k)$ & $\coloneqq Q\, p^0,\dots,p^{k-1} \ldot \qbftransformer(\varphi, i, k)$ for $Q \in \set{\exists,\forall}$ \enspace.
\end{tabular}\smallskip\\
\begin{corollary} \label{thm:co-safety-qbf}
Given a QPTL formula $\Phi = \mathcal{Q} \ldot \varphi$ where $\varphi$ is a syntactically co-safe LTL formula.
If $\qbftransformer(\Phi,0,k)$ is satisfiable for some $k > 0$ then $\Phi$ is satisfiable.
\end{corollary}\noindent
The QBF encoding of CL$_\exists$ formula $\Phi$ and functions $K_1 \dots K_n : \mathcal{C} \rightarrow \mathbb{N}$ is
\begin{align} \label{eq:node_failure_transform_qbf}
  \unsatnf^\mathit{QBF}(\Phi,k,K_1,\dots,K_n) \coloneqq
  \header^\mathit{QBF}(\mathcal{S}, K,k) \ldot
  \consistent^\mathit{QBF}(\mathcal{S},K,k) \rightarrow \nonumber \\
  \bigvee_{1 \leq i \leq n}
  \Big( \bigwedge_{\pi \in \branches(\mathcal{C},K_i)} \varphi_{\mathit{path}_i}^\mathit{QBF}(\pi) \Big) \land
  \Big( \bigvee_{\pi \in \branches(\mathcal{C},K_i)} \qbftransformer(\neg\varphi_i(\pi),0,k) \Big) \enspace,
\end{align}
where $K : \mathcal{C} \rightarrow \mathbb{N}$ is defined as $K(c) \coloneqq \max_{1 \leq i \leq n} K_i(c)$ for every $c \in \mathcal{C}$ and $k > 0$.
$\header^\mathit{QBF}$ introduces the variables $x^0, \dots, x^{k-1}$ that correspond to the path variables $x$, but is otherwise identical to the QPTL version.
Furthermore, $\varphi_\mathit{path}^\mathit{QBF}$ uses the equivalence $\bigwedge_{i < k} ( s^i \leftrightarrow c^i )$ for $\eclalws (c \leftrightarrow v)$.
The consistency condition $\consistent^\mathit{QBF}(\mathcal{S}, K, k)$ in QBF is defined as the conjunction of
\begin{equation*}
  \bigwedge_{(\pi,\pi') \in \branches(\deps(s),K)^2} \Big( \bigwedge_{i < k} \big( s_\pi^i \leftrightarrow s_{\pi'}^i \lor \bigvee_{j < i} \;\bigvee_{c \in \Scope(s)} c_\pi^j \nleftrightarrow c_{\pi'}^j \big) \Big)
\end{equation*}
for every $s \in \mathcal{S}$.
This formula ensures different reactions of a strategy variable on two branches is based on different prior valuations to the dependencies (cf.~QPTL consistency condition~\eqref{eq:consistency_conditon_qptl}).
In this simple translation, one cause of high complexity is due to the consistency conditions between the strategy variables across different paths.
However, most of these variables are not used for the counterexample itself but appear only in the consistency condition.
One optimization removes these unnecessary variables from the encoding.
Therefore, we collect all strategy variables and (when possible) their temporal occurrence from the LTL specification.
For every used strategy variable we build the \emph{dependency graph} that contains all variables which can influence the outcome of the strategy.
In the last step, we remove all variables that are not contained in any dependency graph.
This optimization is depicted in Fig.~\ref{fig:qbf_encoding_optimization_dependency_graph}.

Let us consider again the pipeline example given by the CL$_\exists$ formula (\ref{eq:pipeline_architecture}) $\eclexists{\set{a}}{x} \eclexists{\set{b}}{y} \eclalws (b = x) \rightarrow \eclalws (\eclnext y \leftrightarrow a)$.
The QBF encoding with unrolling depth of $2$ is
\begin{align}
  \exists a_1^0, a_1^1, a_2^0, a_2^1 \ldot \forall x_1^0, x_1^1, x_2^0, x_2^1 \ldot \forall y_1^0, y_1^1, y_2^0, y_2^1 \ldot \exists b_1^0, b_1^1, b_2^0, b_2^1 \ldot \nonumber \\
  (x_1^0 \leftrightarrow x_2^0) \land (x_1^1 \leftrightarrow x_2^1 \lor a_1^0 \nleftrightarrow a_2^0) \land \nonumber \\
  (y_1^0 \leftrightarrow y_2^0) \land (y_1^1 \leftrightarrow y_2^1 \lor b_1^0 \nleftrightarrow b_2^0) \rightarrow \nonumber \\
  \bigwedge_{i \in \set{1,2}} \big( (b_i^0 \leftrightarrow x_i^0) \land (b_i^1 \leftrightarrow x_i^1) \big) \land \bigvee_{i \in \set{1,2}} ( y_i^1 \nleftrightarrow a_i^0 ) \enspace.
\end{align}

\begin{theorem}[Correctness $\unsatnf^\mathit{QBF}$] \label{thm:correctness_qbf_encoding}
  Given a CL$_\exists$ formula $\Phi = \qheader_\exists \ldot \bigwedge_{1 \leq i \leq n} \big( \varphi_{\text{path}_i} \allowbreak \rightarrow \varphi_i \big)$ over coordination variables $\mathcal{C}$ and strategy variables $\mathcal{S}$.
  Let $\varphi_i$ be a syntactically safe LTL formula for each $1 \leq i \leq n$.
  $\Phi$ is unsatisfiable if there exists functions $K_1 \dots K_n : \mathcal{C} \rightarrow \mathbb{N}$ such that the QBF query $\unsatnf^\text{QBF}(\Phi,k,K_1,\dots,K_n)$ is satisfiable for some $k > 0$.
\end{theorem}
\proof
The encoding is a special case of Theorem~\ref{thm:correctness_ws1s_encoding}.
If the Boolean formula that is a finite $k$-unrolling of the LTL formula is satisfied against a finite sequence of length $k$, i.e., we found a counterexample within the first $k$ steps, then it is also satisfied against any (possibly infinite) strategy.
\qed

\section{Experimental Results}

We have carried out our experiments on a 2.6\,GHz Opteron system.
For solving the QBF instances, we used a combination of the QBF preprocessor Bloqqer~\cite{DBLP:conf/cade/BiereLS11} in version 031 and the QBF solver DepQBF~\cite{DBLP:journals/jsat/LonsingB10} in version 3.0.3.
For solving the WS1S instances, we used Mona~\cite{DBLP:conf/tacas/HenriksenJJKPRS95} in version 1.4-15.

\paragraph{\bf Byzantine Generals' Problem.}

\begin{table}[b]
\caption[]{Result of the \emph{Byzantine Generals' Problem} example}\medskip
\label{tbl:bgp_example}
\centering
\begin{tabular}{l|rrrrr}
Bound         & $(0,0,0,0)$   & $(1,0,0,0)$   & $(1,1,0,0)$   & $(1,1,1,0)$   & $(1,1,1,1)$ \\[2pt]\hline
Result        & Unsatisfiable & Unsatisfiable & Unsatisfiable & Unsatisfiable & Satisfiable \\
\#\,Clauses   & 57            & 228           & 2286          & 2904          & 3522 \\
\#\,Variables & 44            & 143           & 1095          & 1375          & 1655 \\
Time\,(s)     & 0.09          & 0.16          & 0.98          & 1.01          & 23.61 \\ \hline
\end{tabular}\\ {\smallskip}
\begin{minipage}[t]{0.915\textwidth}
{\small The table shows the solving time of the Byzantine Generals' Problem $\Phi_\text{bgp}$ (Equation~\ref{eq:byzantine_generals}) using the QBF encoding with a fixed length of $3$ unrollings.
The QBF queries are solved using a combination of Bloqqer~031 and DepQBF~3.0.3.}
\end{minipage}
\end{table}

Table~\ref{tbl:bgp_example} demonstrates that the Byzantine Generals' Problem remains, despite the optimizations described above, a nontrivial combinatorial problem: we need to find a suitable set of paths for every possible combination of the strategies of the generals.
The bound given in the first row reads as follows: The first component is the number of branchings for the input variable $v$ in all three architectures.
The last three components state the number of branchings for the outputs of the faulty nodes in their respective architectures.
For example, bound $(1,1,0,0)$ means that we have two branches for $v$, $c_{12}$, and $c_{13}$, while we have only one branch for $c_{23}$ and $c_{32}$.
More precisely, starting from constant zero functions $K_1,K_2,K_3$, the bound $(1,1,0,0)$ sets $K_1(v)=K_2(v)=K_3(v)=K_1(c_{12})=K_1(c_{13}) = 1$ and $K_2(c_{23}) = K_3(c_{32}) = 0$.
To prove the unrealizability, we need one branching for the input $v$ and one branching for every coordination variable that serves as a shared variable for a faulty node, i.e., the bound $(1,1,1,1)$.
The number of branches and thereby the formula size grows exponentially with the number of branchings for the input variables.

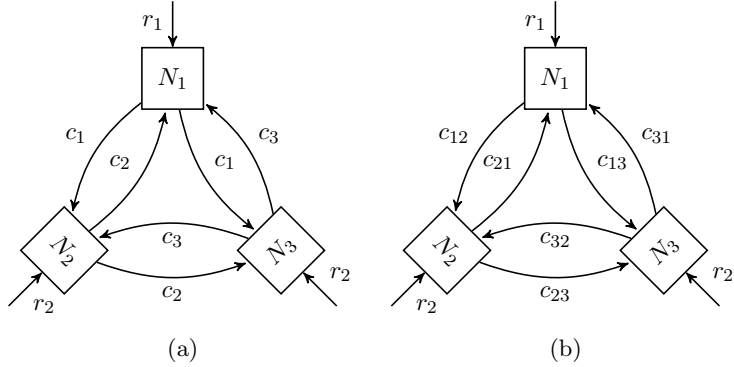
\begin{figure}[t]
\subfigure[]{
  \begin{tikzpicture}[->,>=stealth',shorten >=1pt,auto,node distance=2cm,semithick,scale=0.85,transform shape]

  \tikzstyle{every state}=[shape=rectangle]

  \coordinate (center) at (0,0);
  
  \node[state,above=of center] (n1) {$N_1$};
  \node[state,below left=0.2 and 1 of center,rotate=-45] (n2) {$N_2$};
  \node[state,below right=0.2 and 1 of center,rotate=45] (n3) {$N_3$};
  
  \draw (n1) edge[bend right=20] node[swap] {$c_1$} (n2)
        (n2) edge[bend right=20] node {$c_2$} (n1)
        (n1) edge[bend right=20] node {$c_1$} (n3)
        (n3) edge[bend right=20] node[swap] {$c_3$} (n1)
        (n2) edge[bend right=20] node[swap] {$c_2$} (n3)
        (n3) edge[bend right=20] node {$c_3$} (n2)
        ;
  
  \draw (n1) edge[<-] node {$r_1$} +(0,1.25)
        (n2) edge[<-] node {$r_2$} +(-0.9,-0.9)
        (n3) edge[<-] node {$r_2$} +(0.9,-0.9)
        ;

\end{tikzpicture}
  \label{fig:byzantine_firing_squad_architecture}
}%
\subfigure[]{
  \begin{tikzpicture}[->,>=stealth',shorten >=1pt,auto,node distance=2cm,semithick,scale=0.85,transform shape]

  \tikzstyle{every state}=[shape=rectangle]

  \coordinate (center) at (0,0);
  
  \node[state,above=of center] (n1) {$N_1$};
  \node[state,below left=0.2 and 1 of center,rotate=-45] (n2) {$N_2$};
  \node[state,below right=0.2 and 1 of center,rotate=45] (n3) {$N_3$};
  
  \draw (n1) edge[bend right=20] node[swap] {$c_{12}$} (n2)
        (n2) edge[bend right=20] node {$c_{21}$} (n1)
        (n1) edge[bend right=20] node {$c_{13}$} (n3)
        (n3) edge[bend right=20] node[swap] {$c_{31}$} (n1)
        (n2) edge[bend right=20] node[swap] {$c_{23}$} (n3)
        (n3) edge[bend right=20] node {$c_{32}$} (n2)
        ;

  \draw (n1) edge[<-] node {$r_1$} +(0,1.25)
        (n2) edge[<-] node {$r_2$} +(-0.9,-0.9)
        (n3) edge[<-] node {$r_2$} +(0.9,-0.9)
        ;

\end{tikzpicture}
  \label{fig:cap_theorem_architecture}
}%
\caption[]{Broadcasting~\subref{fig:byzantine_firing_squad_architecture} and non-broadcasting~\subref{fig:cap_theorem_architecture} communication topologies. (We omit the environment process to increase readability.)}
\label{fig:communication_topologies}
\end{figure}

\paragraph{\bf Byzantine Firing Squad Problem.}
\begin{table}[h]
\caption[]{Result of the \emph{Byzantine Firing Squad Problem} example}
\label{tbl:bfsp}
\centering
\begin{tabular}{llllll}
\hline\noalign{\smallskip}
Instance\hspace{10pt} & Result\hspace{40pt} & \#\,Clauses\hspace{10pt} & \#\,Variables\hspace{10pt} & Time\,(s) \\ \noalign{\smallskip}
\hline
\noalign{\smallskip}
bfsp\_3  & Satisfiable & 803      & 435      & 0.21 \\
bfsp\_5  & Satisfiable & 1773     & 967      & 0.43 \\
bfsp\_10 & Satisfiable & 7348     & 3942     & 3.07 \\
bfsp\_20 & Satisfiable & 42\,198  & 22\,042  & 15.58 \\
bfsp\_30 & Satisfiable & 128\,648 & 66\,342  & 66.29 \\
bfsp\_40 & Satisfiable & 290\,698 & 148\,842 & 190.90 \\
bfsp\_50 & Satisfiable & 552\,348 & 281\,542 & 396.20 \\ \hline
\end{tabular}\\ {\smallskip}
\begin{minipage}[t]{0.74\textwidth}
{\small The table shows the solving time of the Byzantine Firing Squad Problem $\Phi_\text{bfsp}$ (Equation~\ref{eq:byzantine_firing_squad}) using the QBF encoding with a fixed length of $3$ unrollings.
The QBF queries are solved using a combination of Bloqqer~031 and DepQBF~3.0.3.}
\end{minipage}
\end{table}
We consider another classical consensus problem, the \emph{Byzantine Firing Squad Probem}~\cite{DBLP:conf/podc/FischerLM85}.
In this setting, we have a set of synchronous processes, each having a distinct input and a broadcasting mechanism for communication with the other processes, depicted in Fig.~\ref{fig:byzantine_firing_squad_architecture}.
The goal of these processes is to synchronously enter a \emph{fire} state whenever the input was initially given at one process.
As before, we consider architectures with up to one faulty process, but in contrast to the Byzantine Generals' Problem, we require that this specification holds only if all processes are correct.
Furthermore, we require that the non-faulty processes act \emph{uniformly}, i.e., if a non-faulty process enters the firing state, all other non-faulty processes enter the firing state.
It was shown in~\cite{DBLP:conf/podc/FischerLM85} that there is no protocol that ensures correct behavior in the presence of faults.
The encoding $\Phi_\text{bfsp}$ for three nodes is given by the CL$_\exists$ formula
\begin{align} \label{eq:byzantine_firing_squad}
  &\eclexists{\set{\text{req}_1}}{\text{bcast}_1} \eclexists{\set{\text{req}_1, \text{chan}_{21}, \text{chan}_{31}}}{\text{out}_1} \nonumber\\
  &\eclexists{\set{\text{req}_2}}{\text{bcast}_2} \eclexists{\set{\text{req}_2, \text{chan}_{12}, \text{chan}_{32}}}{\text{out}_{2}} \nonumber\\
  &\eclexists{\set{\text{req}_3}}{\text{bcast}_3} \eclexists{\set{\text{req}_3, \text{chan}_{13}, \text{chan}_{23}}}{\text{out}_{3}} \nonumber\\
  & \left( \eclalws  \operational_{2,3}  \rightarrow \consistent_{2,3} \right) \land {} \nonumber\\
  & \left( \eclalws  \operational_{1,3}  \rightarrow \consistent_{1,3} \right) \land {} \nonumber\\
  & \left( \eclalws  \operational_{1,2}  \rightarrow \consistent_{1,2} \right) \land {} \nonumber\\
  & \left( \eclalws  \operational_{1,2,3}  \rightarrow \consistent_{1,2,3} \land \uniform_{1,2,3} \right) \enspace,
\end{align}\noindent
where
\begin{align}
  \consistent_N &{} \coloneqq \eclalws \bigwedge_{i<j \in N^2} \big( \text{out}_i = \text{out}_j \big) \text{ and} \nonumber \\
  \uniform_N &{} \coloneqq \left( \bigvee_{i \in N} \text{req}_i \right) \rightarrow \eclnnext{3} \left( \bigvee_{i \in N} \text{out}_i \right) \enspace. \label{eq:bfsp_fixed_uniform}
\end{align}
Table~\ref{tbl:bfsp} shows that our method can detect conflicts quickly even though the CL$_\exists$ encoding $\Phi_\text{bfsp}$ grows quadratically in the number of nodes.

\paragraph{\bf CAP Theorem.}
\noindent
The CAP Theorem due to Brewer~\cite{DBLP:conf/podc/Brewer00} states that it is impossible to design a distributed system that provides Consistency, Availability, and Partition tolerance (CAP) simultaneously.
For the encoding in CL$_\exists$, we assume there is a fixed number $n$ of nodes, that every node implements the same service, and that there are direct communication links between all nodes, depicted in Fig.~\ref{fig:cap_theorem_architecture}.
We use the variables $\text{req}_i$ and $\text{out}_i$ to denote input and output of node $i$, respectively.
The consistency and availability requirements are encoded as the LTL formulas $\bigwedge_{1 \leq i < n} (\text{out}_i \leftrightarrow \text{out}_{i+1})$ and $(\bigvee_{1 \leq i \leq n} \text{req}_i) \leftrightarrow (\eclnnext{2} \bigvee_{1 \leq i \leq n} \text{out}_i)$.
The partition tolerance is modeled in a way that there is always at most one node partitioned from the rest of the system, i.e., we have $n$ different architectures and in every architecture all communication links to one node are faulty.
For two nodes, we get the CL$_\exists$ formula $\Phi_\text{cap}$
\begin{align} \label{eq:cap_theorem}
  &\eclexists{\set{\text{req}_1}}{\text{com}_1} \eclexists{\set{\text{req}_1, \text{chan}_2}}{\text{out}_1} \eclexists{\set{\text{req}_2}}{\text{com}_2} \eclexists{\set{\text{req}_2, \text{chan}_1}}{\text{out}_2} \nonumber\\
  & ( \eclalws ( \text{chan}_1 = \text{com}_1 ) \rightarrow \eclalws ( (\text{out}_1 = \text{out}_2) \land ( (\text{req}_1 \lor \text{req}_2) \leftrightarrow \eclnnext{3} (\text{out}_1 \lor \text{out}_2) ) ) ) \land {} \nonumber\\
  & ( \eclalws ( \text{chan}_2 = \text{com}_2 ) \rightarrow \eclalws ( (\text{out}_1 = \text{out}_2) \land ( (\text{req}_1 \lor \text{req}_2) \leftrightarrow \eclnnext{3}(\text{out}_1 \lor \text{out}_2) ) ) ) \enspace .
\end{align}\noindent%
Table~\ref{tbl:cap_theorem} shows that our method is able to find conflicts in a specification with an architecture up to 50 nodes within reasonable time.
When we drop either Consistency, Availability, or Partition tolerance, the corresponding instances (AP, CP, and CA) become satisfiable.
Hence, our tool does not find counterexamples in these cases.

\begin{table}[b]
\caption[]{Result of the \emph{CAP Theorem} example}
\label{tbl:cap_theorem}
\centering
\begin{tabular}{llllll}
\hline\noalign{\smallskip}
Instance\hspace{10pt} & Result\hspace{40pt} & \#\,Clauses\hspace{10pt} & \#\,Variables\hspace{10pt} & Time\,(s) \\ \noalign{\smallskip}
\hline
\noalign{\smallskip}
ap\_2 & Unsatisfiable & 1232    & 619     & 0.41 \\
ca\_2 & Unsatisfiable & 1408    & 763     & 1.11 \\
cp\_2 & Unsatisfiable & 48      & 42      & 0.09 \\
cap\_2 & Satisfiable  & 110     & 84      & 0.11 \\
cap\_5 & Satisfiable  & 665     & 426     & 0.28 \\
cap\_10 & Satisfiable & 2590    & 1556    & 1.07 \\
cap\_25 & Satisfiable & 15\,865 & 9146    & 20.60 \\
cap\_50 & Satisfiable & 62\,990 & 35\,796 & 781.16 \\ \hline
\end{tabular}\\ {\smallskip}
\begin{minipage}[t]{0.74\textwidth}
{\small The table shows the solving time of the CAP Theorem $\Phi_\text{cap}$ (Equation~\ref{eq:cap_theorem}) using the QBF encoding with a fixed length of $3$ unrollings.
The QBF queries are solved using a combination of Bloqqer~031 and DepQBF~3.0.3.}
\end{minipage}
\end{table}

\paragraph{\bf Discussion.}

Table~\ref{tbl:comparison_encodings} shows a comparison of the the different encodings that we have presented in the paper.
There does not exist an algorithm that decides whether a given CL$_\exists$ formula is unsatisfiable.
We presented an approach that bounds the number of paths and used an encoding to QPTL\@.
The reason for incompleteness was shown in~Section~\ref{sec:completeness}: for liveness properties one may need infinitely many paths to show unsatisfiability.
Our encoding in WS1S (Mona) loses the ability to find counterexample paths of infinite length, e.g., the CL$_\exists$ formula
\begin{equation} \label{eq:example_infinite_long}
  \eclexists{\emptyset}{y} \eclevtlalws (\eclnext y \leftrightarrow x)  
\end{equation}
with free coordination variable $x$ is unsatisfiable and two paths that are infinitely often different are sufficient to prove it.
The QPTL encoding is capable of finding these paths while neither the WS1S, nor the QBF, encoding is applicable.
However, Mona failed to solve larger instances from Tables~\ref{tbl:bgp_example} to~\ref{tbl:cap_theorem}, especially it could not handle the encoding of the Byzantine Generals' Problem.
For the translation to QBF we bounded the length of the paths to $k$ where $k$ is an additional parameter.
With this encoding we approximate the reactive behavior of our system by a finite prefix.
Despite this restriction, we could prove unsatisfiability for many interesting specifications from literature.
In some instances, the WS1S approximation provides stronger unrealizability guarantees.
For example in the Byzantine Firing Squad Problem, the WS1S encoding is able to prove that any \emph{finite} uniform reaction
\begin{equation} \label{eq:bfsp_finite_uniform}
  \uniform_N \coloneqq \left( \bigvee_{i \in N} \text{req}_i \right) \rightarrow \eclevtl \left( \bigvee_{i \in N} \text{out}_i \right)
\end{equation}
leads to an unrealizable instance (cf.\ Table~\ref{tbl:comparison_encodings}), while in the QBF encoding, we have to bound the uniform reaction~(\ref{eq:bfsp_fixed_uniform}) to a fixed length.
In practice, one would first use the QBF approximation in order to find ``cheap'' counterexamples.
After hitting the number of paths that the QBF solver can no longer handle within reasonable time, one proceeds with more costly approximations like the WS1S encoding.
\begin{table}
\caption[]{Comparison of different encodings}
\label{tbl:comparison_encodings}
\centering
\begin{tabular}{lcccccccc}
\hline\noalign{\smallskip}
 & independent & pipeline & join & bgp & bfsp & cap & infinite  & bfsp-finite \\
 & (\ref{eq:undecidable_architecture}) & (\ref{eq:pipeline_architecture}) & (\ref{eq:node_failure_example}) & (\ref{eq:byzantine_generals}) & (\ref{eq:byzantine_firing_squad}) & (\ref{eq:cap_theorem}) & (\ref{eq:example_infinite_long}) & (\ref{eq:byzantine_firing_squad}),(\ref{eq:bfsp_finite_uniform}) \\
\noalign{\smallskip}\hline\noalign{\smallskip}
QPTL & \checkmark & (\checkmark) & (\checkmark) & (\checkmark) & (\checkmark)   & (\checkmark)   & \checkmark & (\checkmark) \\
WS1S & \checkmark & \checkmark   & \checkmark   & (\checkmark) & \checkmark (5) & \checkmark (6) & -          & \checkmark (5) \\
QBF  & \checkmark & \checkmark   & \checkmark   & \checkmark   & \checkmark     & \checkmark     & -          & - \\ \hline
\end{tabular}\\ {\smallskip}
\begin{minipage}[t]{0.873\textwidth}
{\small The table compares the expressiveness of the encodings and the experimental verifiability on the examples presented in this paper.
The $(\checkmark)$ symbol denotes that the tool failed to solve the query, the number behind the checkmark indicates up to how many nodes the query was successful.
We used GOAL, Mona~1.4-15, and a combination of Bloqqer~031 and DepQBF~3.0.3 for the QPTL, WS1S, and QBF queries, respectively.}
\end{minipage}
\end{table}


\section{Beyond the Byzantine Fault Model}

We formalize the fault-tolerant synthesis problem regarding multiple types of faults, their durations, observability, and the fault-tolerance requirements and provide a uniform encoding into CL$_\exists$.
Faults are modeled by allowing the environment to take over the control of the faulty processes.

\paragraph{\bf Types of Faults.}
We review types of faults that were considered in previous work~\cite{DBLP:conf/ftdc/DolevS86,DBLP:journals/toplas/AttieAE04,DBLP:conf/atva/DimitrovaF09}.
\emph{Stuck-at faults} occur when a process stops reacting on new inputs and keep their last outputs.
On \emph{fail-stop faults} or \emph{crashes} the process halts immediately, i.e., they stop producing any outputs.
While \emph{fail-stop faults} are detectable, \emph{crashes} are not detectable~\cite{DBLP:conf/ftdc/DolevS86}.
\emph{Omission faults} subsume these faults and are characterized as failures on the specified input-output specification.
The most general fault type are the \emph{Byzantine faults}, where a process can deviate arbitrarily from its specification, even in a malicious way.

The duration of a fault can be \emph{permanent}, \emph{intermittent}, or \emph{transient}.
Some failures may be \emph{detectable} or \emph{undetectable} depending on the architecture and the type of fault.

Lastly, the system usually has to fulfill different kind of specifications in the presence of faults and we use the classification of fault-tolerance requirements from~\cite{DBLP:conf/atva/DimitrovaF09}: \emph{Masking tolerance} means that the system has to fulfill both safety and liveness specifications, \emph{non-masking tolerance} allows the system to temporarily violate its safety properties while \emph{fail-safe tolerance} specifies that after a fault, only the safety requirements have to be fulfilled.

\paragraph{\bf Fault-tolerance Specifications.}
For simplicity, we restrict the failures to processes, but it can be easily adapted to failures of individual communication links.
Consider an arbitrary architecture $\mathcal{A} = (P,p_\mathit{env},\{I_p\}_{p \in P}, \{O_p\}_{p \in P})$.
A \emph{fault-tolerance scenario} $F$ is a tuple $(P_f,T,D,\mathit{Obs},\varphi)$ where $P_f \subseteq P$ is a subset of processes that simultaneously suffer from a fault of type $T \in \set{\mathit{stuck\text{-}\hspace{-0.4pt}at}, \emph{fail-stop}, \emph{omission},\emph{Byzantine}}$ and duration $D \in \set{\mathit{permanent}, \mathit{intermittent}, \mathit{transient}}$, $\mathit{Obs} \in \set{\btrue,\bfalse}$ denotes the observability, and $\varphi$ is an LTL formula specifying the fault-tolerance requirements.
Not all combinations of fault types and durations are possible, e.g., stuck-at faults and fail-stop faults are always permanent.
A \emph{fault-tolerance specification} $\mathcal{F}$ is a set of fault-tolerance scenarios.

\paragraph{\bf Fault-tolerant Realizability.}
The distributed fault-tolerant realizability problem is to decide, given an architecture $\mathcal{A}$ and a fault-tolerance specification $\mathcal{F}$, whether there exists a finite state implementation for the processes in $\mathcal{A}$ such that the system satisfies the fault-tolerance specification $\mathcal{F}$.
We formalize this problem by giving a CL$_\exists$ formula $\Phi = \mathcal{Q}_\exists \ldot \bigwedge_{F \in \mathcal{F}} \big( \varphi_{\mathit{path}_F} \rightarrow \varphi_F \big)$ that models the description of the fault-tolerance scenarios given above.

Consider an arbitrary architecture $\mathcal{A} = (P,p_\mathit{env},\{I_p\}_{p \in P}, \{O_p\}_{p \in P})$ and a fault-tolerance specification $\mathcal{F}$.
Using Theorem~\ref{thm:ecl_encoding_distributed_synthesis} we get the CL$_\exists$ encoding $\Psi = \mathcal{Q}_\exists \ldot \varphi_\mathit{path} \rightarrow \varphi$ of the distributed realizability problem for architecture $\mathcal{A}$ and an arbitrary LTL formula $\varphi$.
For every fault-tolerance scenario $F \in \mathcal{F}$, we introduce a new architecture that modifies the input-output specifications for processes affected by $F$ according to the scenario.
We denote the resulting path specification by the LTL formula $\varphi_{\mathit{path}_F}$ and the fault-tolerance requirement for $F$ as $\varphi_F$.
The resulting CL$_\exists$ formula contains a conjunct over all scenarios $\mathcal{F}$:
\begin{equation*}
  \Phi = \mathcal{Q}_\exists \ldot \bigwedge_{F \in \mathcal{F}} \big( \varphi_{\mathit{path}_F} \rightarrow \varphi_F \big)
\end{equation*}

We now define the path specifications $\varphi_{\mathit{path}_F}$.
Let $F = (P_f,T,D,\mathit{Obs},\varphi_F)$ be a fault-tolerance scenario.
For every process $p \in P_f$ we introduce a fresh coordination variable $\mathit{fault}^F_p$ that immediately signals whether a fault occurred at process $p$ in the last step.
If the faults are observable ($\mathit{Obs} = \btrue$) we extend the scope of every strategy variable by $\set{\mathit{fault}^F_p \mid p \in P_f}$\footnote{These signals are called \emph{fault-notification variables} in~\cite{DBLP:conf/atva/DimitrovaF09} as only detectable faults were considered.}.

Let $\varphi_{\mathit{path}_F} = \varphi_\mathit{path} = \bigwedge_{v \in I \cap O} \eclalws (c_v = s_v)$.
We remove every conjunct $\eclalws (c_v = s_v)$ from $\varphi_{\mathit{path}_F}$ where $v \in O_{P_f}$ is the output of a faulty process.
For every $v \in I \cap O_{P_f}$ we add the following new constraints to $\varphi_{\mathit{path}_F}$ depending on the type of faults $T$.
For readability, we replace $\mathit{fault}^F_p$ by $\mathit{fault}$, $c_v$ by $i$, and $s_v$ by $o$.
\begin{itemize}
  \item $T = \mathit{stuck\text{-}\hspace{-0.4pt}at}$: $\eclwuntil{\big((\neg\mathit{fault} \land (i = o)\big)}{\big( \eclalws \mathit{fault} \land i \leftrightarrow \eclnext\eclalws i \big)}$.
  \item $T = \mathit{fail\text{-}\hspace{-0.4pt}stop}$: $\eclwuntil{\big(\neg\mathit{fault} \land (i = o)\big)}{\big(\eclalws (\mathit{fault} \land \neg i)\big)}$.
  \item $T = \mathit{omission}$: $\eclalws \big(\mathit{fault} \land \neg i \lor \neg\mathit{fault} \land (i = o)\big)$.
  \item $T = \mathit{Byzantine}$: $\eclalws \big(\mathit{fault} \lor \neg\mathit{fault} \land (i = o)\big)$.
\end{itemize}
Lastly, we add a requirement for every $\mathit{fault}^F_p$ variable regarding the specified duration $D$:
\begin{itemize}
  \item $D = \mathit{permanent}$: $\eclwuntil{\neg \mathit{fault}}{(\eclalws \mathit{fault})}$
  \item $D = \mathit{intermittent}$: $\eclalwsevtl \neg\mathit{fault}$. If we want to have tighter control about the duration of faults we can use the parametric LTL operators~\cite{DBLP:journals/tocl/AlurETP01} to bound the length $n$ of a fault $\ltlalws\ltlevtl_{\leq n}\, \neg\mathit{fault}$, or require that length of a correct behavior is longer than $m$ by formula $\eclalws \big(\mathit{fault} \rightarrow \ecluntil{\mathit{fault}}{ (\eclpuntil{> m}{\neg\mathit{fault}}{\mathit{fault} }) } \big)$.
  \item $D = \mathit{transient}$: $\eclwuntil{\neg\mathit{fault}}{ \big(\eclpuntil{\leq n}{\mathit{fault}}{\eclalws \neg\mathit{fault} }\big) }$ where $n$ is the maximal duration of the fault.
\end{itemize}
We simplify the undetectable permanent fault encoding by removing the $\mathit{fault}$ variables, as these variables are not used in the specification.
Our formulation allows for many types and durations of faults that cannot be captured with a fault-tolerance scenario.
Indeed, it is possible to use arbitrary LTL formulas to specify the type and duration of a fault.
Computing counterexamples for these general types of faults can be integrated into the method described in Section~\ref{sec:computing_counterexamples}.

\section{Conclusion}

We have introduced counterexamples for distributed realizability and shown how to automatically derive counterexamples from given specifications in CL$_\exists$.
We used encodings in QPTL, WS1S, and QBF\@.
In our experiments, the QBF encoding was the most efficient.
Even problems with high combinatorial complexity, such as the Byzantine Generals' Problem, are handled automatically.
Given that QBF solvers are likely to continue to improve in the future, even larger instances should become tractable.
%
In future work, we plan to extend the method to a larger class of \emph{infinite} counterexamples, which will support liveness specifications.
Furthermore, we want to investigate approximation techniques for more general types of faults.

\section*{Acknowledgments}

This work was partially supported by the German Research Foundation (DFG) as part of the Transregional Collaborative Research Center ``Automatic Verification and Analysis of Complex Systems'' (SFB/TR 14 AVACS).
We thank Swen Jacobs for comments on an earlier version of this paper.

\bibliographystyle{alpha}
\bibliography{main}

\end{document}